\documentclass[a4paper,12pt]{article}

\usepackage{bbold}
\usepackage{float}
\usepackage{cancel}
\usepackage{subfig}
\usepackage{amsmath, amssymb, setspace,cite}
\usepackage{slashed}
\usepackage{graphicx}
\usepackage[hidelinks]{hyperref}
\usepackage{color}
\usepackage{soul}
\usepackage{xcolor}
\usepackage{braket}
\usepackage{makecell}
\usepackage{hhline}

\numberwithin{equation}{section}
\vspace{0.6cm}
\date{\today} 

\usepackage{tabularx}
\usepackage{longtable} 
\usepackage{multicol}
\usepackage{multirow}
\usepackage{enumitem}

\renewcommand{\arraystretch}{1.2}

\usepackage{marginnote}
\begin{document}

\def\thefootnote{\fnsymbol{footnote}}

\begin{center}
\Large\bf\boldmath
\vspace*{1.cm} 
Potential of light-cone sum rules without semiglobal quark-hadron duality
\unboldmath
\end{center}
\vspace{0.6cm}

\begin{center}
A.~Carvunis$^{1,}$\footnote{Electronic address: alexandre.carvunis@unito.it}, 
F.~Mahmoudi$^{2,3,4,}$\footnote{Electronic address: nazila@cern.ch}, 
Y. Monceaux$^{2,}$\footnote{Electronic address: y.monceaux@ip2i.in2p3.fr}\\
\vspace{0.6cm}
{\sl $^1$ Dip. di Fisica, Università di Torino and INFN, Sezione di Torino \\
via Giuria 1, 10125 Torino, Italy}\\[0.4cm]
{\sl $^2$Universit\'e Claude Bernard Lyon 1, CNRS/IN2P3, \\
Institut de Physique des 2 Infinis de Lyon, UMR 5822, F-69622, Villeurbanne, France}\\[0.4cm]
{\sl $^3$Theoretical Physics Department, CERN, CH-1211 Geneva 23, Switzerland}\\[0.4cm]
{\sl $^4$Institut Universitaire de France (IUF), 75005 Paris, France}
\end{center}

\renewcommand{\thefootnote}{\arabic{footnote}}
\setcounter{footnote}{0}

\vspace{1.cm}
\begin{abstract}

This work addresses the calculation of local form factors involved in the theoretical predictions of semileptonic $B$-meson decays at low $q^2$.
We present a new approach to the method of QCD light-cone sum rule with $B$-meson light-cone distribution amplitudes. In our strategy, we bypass the semiglobal quark-hadron duality (QHD) approximation which usually contributes an unknown and potentially large systematic error to the prediction of form factors. We trade this improvement for an increased reliance on higher-order contributions in perturbation theory. Unlike the systematic error from semiglobal QHD, truncation errors are assessable and systematically improvable, hence allowing robust predictions of form factors. For the transitions $B\to \pi,\rho,K^{(*)}$, our predictions agree with the literature for all form factors at $q^2=0$. 
\end{abstract}

\newpage

\section{Introduction}
For a decade, numerous deviations from the predictions of the Standard Model (SM) have been measured in $B$-meson decays, particularly in $b \to s \ell^+ \ell^-$ transitions (the so-called anomalies). The most precisely predicted observables were the lepton flavour universality ratios of a $B$ meson decaying to $K^{(*)}$ and a pair of muons compared to the same decay with electrons in the final state. These ratios were known to exhibit a $\sim 25\%$ deficit compared to the SM predictions, which are expected to be close to 1 with high accuracy~\cite{LHCb:2017avl,LHCb:2021trn}. However, the latest measurements by LHCb of these ratios now show good agreement with the SM predictions~\cite{LHCb:2022vje}.

Nevertheless, anomalies persist in other observables related to $B$-meson decays. Branching fractions and angular observables of $B$ decays to final states $M \mu^+ \mu^-$ where $M = K^{(*)}, \phi$, particularly at low di-lepton momenta $q^2$, have been extensively studied at LHCb~\cite{LHCb:2013ghj,LHCb:2014cxe,LHCb:2015tgy,LHCb:2015wdu,LHCb:2015svh,LHCb:2020lmf,LHCb:2020gog,LHCb:2021xxq,LHCb:2021zwz}. Additionally, the CMS collaboration recently measured BR($B^+ \to K^+ \mu^+\mu^-$)~\cite{CMS:2024syx} and the angular distribution of $B^0 \to K^{*0} \mu^+ \mu^-$~\cite{CMS-PAS-BPH-21-002}, the findings of LHCb. These observables appear to be in tension with the SM predictions, with a significance ranging from $2-4 \sigma$ depending on theoretical assumptions. 
It is crucial to note that the SM predictions for these observables are very sensitive to non-perturbative contributions, introducing a notable source of uncertainty, which dominates the overall error of the predictions.
The latter are typically divided into two categories, the local and the non-local contributions to the matrix elements $\bra{M} J^\mu \ket{B}$. The local contributions, expressed in terms of form factors, can be calculated using QCD sum rules on the light cone which we discuss in this work, and in lattice QCD. The non-local contributions present a greater challenge to quantify and have been suggested as potential sources of the disparities between experimental data and theoretical predictions. However, using unitarity bounds, it was shown in \cite{Bobeth:2017vxj, Gubernari:2020eft} that these contributions do not explain the $B$ anomalies. Moreover, it is possible to argue that the discrepancies arising from non-local effects are expected to have a $q^2$ and helicity-dependent behaviour. Looking at the $q^2$ dependence of putative new physics contributions to local operators in the weak effective theory, it has been shown that experimental data are consistent with purely $q^2$-independent local contributions (see e.g. Refs.~\cite{Jager:2012uw,Jager:2014rwa,Ciuchini:2015qxb,Capdevila:2017ert,Chobanova:2017ghn,Arbey:2018ics,Ciuchini:2018anp,Hurth:2020rzx,Bordone:2024hui}). While the significance of the anomalies depends on the employed methods and assumptions for the non-local contributions~\cite{Mahmoudi:2024zna}, we do not delve deeper into this aspect in the present paper. Instead, our focus in this work is on studying the local contributions.

In semileptonic $B$ decays, the measured discrepancies occur at low $q^2$ where most hadronic form factors are difficult to compute with lattice QCD. For $B$ to light meson decays, the only lattice results available at $q^2 = 0$ are for $f_{+,T}^{B \to K}$ \cite{Parrott:2022rgu}. The QCD light-cone sum rules (LCSR) techniques are typically employed in this regime \cite{Khodjamirian:1997lay, Braun:1999dp, Ball:2004rg, Ball:2004ye, Khodjamirian:2005ea, Khodjamirian:2006st, Duplancic:2008ix, Duplancic:2008tk,Khodjamirian:2010vf, Bharucha:2012wy, Wang:2015vgv, Rusov:2017chr,Khodjamirian:2017fxg, Lu:2018cfc, Cui:2022zwm,Monceaux:2023byy, Gubernari:2023rfu, Khodjamirian:2023wol}, albeit with a systematic error stemming from the use of both global and semiglobal quark-hadron duality (QHD) approximation. The magnitude of this error remains unknown.

In this paper, we propose a strategy to circumvent the reliance on semiglobal quark-hadron duality and trade the associated unknown systematic error for an increased yet quantifiable and improvable error coming from the truncation of the perturbative QCD expansion and light-cone operator-product expansion (LCOPE).
This strategy relies on the convergence of a sum rule which we derive below. With a limited knowledge of the twist expansion of the distribution amplitudes used in the LCSR and the radiative corrections, convergence may not be reached. We show that in such a case and under certain assumptions, this strategy can be used to derive upper limits on form factors. 
The knowledge of upper limits on form factors is particularly relevant in the current context of the $B$ anomalies since the $b \to s \mu \mu$ branching ratios are experimentally suppressed with respect to current SM predictions, and smaller SM form factors could account for such discrepancies. 

This article is structured as follows: we start by briefly summarising light-cone sum rules with $B$-meson light-cone distribution amplitudes (LCDAs) as established in the literature in Section 2. Our approach to the LCSRs without semiglobal quark-hadron duality is then introduced in Section 3, and the corresponding numerical results are presented in Section 4. Section 5 provides our conclusions.  

\section{Light-cone sum rules with \textit{\textbf{B}}-meson LCDAs}
Light-cone sum rules applied to the calculation of form factors in $B$ decays have been pioneered in \cite{Khodjamirian:1997lay, Braun:1999dp} using light meson distribution amplitudes. Later, LCSRs using $B$-meson LCDAs \cite{Grozin:1996pq, Beneke:2000wa} were developed in \cite{Khodjamirian:2005ea, DeFazio:2005dx}. Further progress has since then been made on $B$-meson LCDAs  \cite{Kawamura:2001bp, Kawamura:2002mg, Lange:2003ff, DeFazio:2007hw ,Kawamura:2010tj, Bell:2013tfa, Feldmann:2014ika, Braun:2014owa, Braun:2015pha, Braun:2017liq, Braun:2019wyx, Feldmann:2022uok}.
In this section, we introduce our notations and the tools needed for deriving the sum rule.

\subsection{Establishing the sum rule}
In order to compute the form factors of a given $B \to M$ process, the fundamental object for LCSRs with $B$-meson LCDAs is the $B$ meson to vacuum correlation function~\cite{Khodjamirian:2005ea, Khodjamirian:2006st, Gubernari:2018wyi}:
\begin{equation}\label{Correlation_function}
    \Pi^{\mu\nu}(q, k) = i \int d^4x e^{ik.x}\bra{0}T \left\{ J^\nu_{int}(x)J^\mu_{weak}(0) \right\} \ket{\bar{B}(p_B = q+k)}\,,
\end{equation}
where $p_B$ is the 4-momentum of the on-shell $B$-meson and $q$ is the momentum transfer. $J^\mu_{weak}$ is a $b \to$ light transition current and $J^\nu_{int}$ is the interpolating current. In Table \ref{tab:currents}, we list the correspondence between these currents and the associated hadronic processes and form factors. $x$ is the space-time separation between the two interaction points associated to the two currents. From analyticity, for a negative $k^2$ the $B$ meson to vacuum correlation function can be written as:
\begin{equation}\label{analyticity}
    \Pi^{\mu \nu}(q, k) = \frac{1}{\pi}\int_{t_{min}}^\infty ds \frac{\text{Im} \Pi^{\mu \nu}(q,s)}{s-k^2}\,, 
\end{equation}
where $t_{min}$ is below every hadronic threshold. The imaginary part of the correlation function can be expressed using a unitarity relation, obtained by inserting a complete set of hadronic states between the two currents:
\begin{equation}
    2  \text{Im}\Pi^{\mu \nu}(q^2,k^2) = 2 \pi \delta(k^2 - m_M^2) \bra{0}J_{\text{int}}^\nu\ket{M(k)}\bra{M(k)}J_{\text{weak}}^\mu \ket{\bar{B}(p_B)} + 2 \pi \rho^{\mu \nu} (q,k)\,,
\end{equation}
where $\rho^{\mu \nu}$ stands for the spectral density of the excited and continuum states, and $M$ is the lightest meson of the aforementioned set of hadronic states, whose contribution has been singled out. This yields the following relation:
\begin{equation}
    \Pi^{\mu \nu}(q,k)= \frac{\bra{0}J^\nu_{int}\ket{M(k)}\bra{M(k)}J^\mu_{weak}\ket{\bar{B}(q+k)}}{m_M^2-k^2}+\int_{s_{cont}}^{+ \infty} ds \frac{\rho^{\mu \nu}(q,s)}{s-k^2}\,,
\end{equation}
where $s_{cont}$ is the threshold of the lowest continuum or excited state. 
 
The $M$ to vacuum matrix element on the rhs can be expressed with the light meson decay constant:
\begin{equation}
\begin{aligned}
    &\bra{0}\bar{q}_2\gamma^\nu \gamma_5 q_1 \ket{P(k)} = i k^\nu f_P\,,\\
    &\bra{0}\bar{q}_2\gamma^\nu q_1 \ket{V(k, \eta)} =  \eta^\nu m_V f_V\,,
\end{aligned}
\end{equation} 
where $q_1$ and $q_2$ are the constituents of the light meson. \\ 
The $B \to M$ matrix elements are linear combinations of the hadronic form factors we wish to compute. In Appendix \ref{appendix:FFdef} we introduce our definition of the form factors, which is the same as~\cite{Gubernari:2018wyi}. The correlation function \eqref{Correlation_function} can be decomposed as a sum of scalar functions times Lorentz structures $\Pi^{\mu \nu}(q^2,k^2)= \sum_F \Gamma^{\mu \nu}_F \Pi_F(q^2,k^2)$. By identifying the Lorentz structures $\Gamma^{\mu \nu}_F$ defined in Table \ref{tab:currents}, one can extract a scalar relation for each form factor which takes the form:

\begin{equation} \label{eq:Pi_p0}
    \Pi_{F}(q^2,k^2) = Y_F \frac{F(q^2)}{m_M^2 - k^2} + \int_{s_{cont}}^\infty \frac{\rho_F(q^2,s)}{s-k^2} \,,
\end{equation}
where $Y_F$ are also defined in Table \ref{tab:currents}. 

\begin{table}[H]
    \renewcommand{\arraystretch}{1.5}
    \centering
    \begin{tabular}{|c|c|c|c|c|c|}
         \hline
         Process & $J_{int}^{\nu}$ & $J_{weak}^{\mu}$ & $\Gamma_F^{\mu \nu}$ & $Y_F$ & Form factor \\ \hline \hline
         \multirow{ 2 }{*}{$\bar{B}^0 \to \pi^+$} & \multirow{ 2 }{*}{$\bar{d} \gamma^\nu \gamma_5 u$} 
         & $\bar{u} \gamma^\mu h_v$ & $k^\mu k^\nu$  & $2 i f_\pi$ & $f_+^{B \to \pi}$\\
         && $\bar{u} \sigma^{\mu\{q\}} h_v$ & $q^\mu k^\nu$ & $\frac{(m_B^2 - m_\pi^2 - q^2)}{m_B + m_\pi} f_\pi$ & $f_T^{B \to \pi}$ \\ \hline
         \multirow{ 2 }{*}{$\bar{B}^0 \to \bar{K}^0$} & \multirow{ 2 }{*}{$\bar{d} \gamma^\nu \gamma_5 s$} 
          & $\bar{s} \gamma^\mu h_v$ & $k^\mu k^\nu$  & $2 i f_K$ & $f_+^{B \to K}$\\
         && $\bar{s} \sigma^{\mu\{q\}} h_v$ & $q^\mu k^\nu$ & $\frac{(m_B^2 - m_K^2 - q^2)}{m_B + m_K} f_K$ & $f_T^{B \to K}$ \\ \hline
          \multirow{ 2 }{*}{$\bar{B}^0 \to D^+$} & \multirow{ 2 }{*}{$\bar{d} \gamma^\nu \gamma_5 c$} 
          & $\bar{c} \gamma^\mu h_v$ & $k^\mu k^\nu$  & $2 i f_D$ & $f_+^{B \to D}$\\
         && $\bar{c} \sigma^{\mu\{q\}} h_v$ & $q^\mu k^\nu$ & $\frac{(m_B^2 - m_D^2 - q^2)}{m_B + m_D} f_D$ & $f_T^{B \to D}$ \\ \hline
         \multirow{ 5 }{*}{$\bar{B}^0 \to \rho^+$}
         &\multirow{ 5 }{*}{$\bar{d} \gamma^\nu u$}
         & $\bar{u} \gamma^\mu h_v$ & $\varepsilon^{\mu \nu \{k q \}}$ & $\frac{2 m_\rho f_\rho}{m_B + m_\rho}$ & $V^{B \to \rho}$ \\
         && $\bar{u} \gamma^\mu \gamma_5 h_v$  & $g^{\mu \nu}$ & $-i m_\rho f_B (m_B + m_\rho)$ & $A_1^{B \to \rho}$ \\
         && $\bar{u} \gamma^\mu \gamma_5 h_v$  & $k^\mu q^\nu$ & $\frac{2 i m_\rho f_\rho}{m_B + m_\rho}$ & $A_2^{B \to \rho}$ \\
         && $\bar{u} \sigma^{\mu\{q\}} h_v$ &  $\varepsilon^{\mu \nu \{k q \}}$ & $2 i m_\rho f_\rho$ & $T_1^{B \to \rho}$ \\
         && $\bar{u} \sigma^{\mu\{q\}} \gamma_5 h_v$ & $q^\mu q^\nu$ & $2 m_\rho f_\rho$ & $T_{23B}^{B \to \rho}$ \\  \hline
         \multirow{ 5 }{*}{$\bar{B}^0 \to \bar{K}^{*0}$}
         &\multirow{ 5 }{*}{$\bar{d} \gamma^\nu s$}
         & $\bar{s} \gamma^\mu h_v$ & $\varepsilon^{\mu \nu \{k q \}}$ & $\frac{2 m_{K^*} f_{K^*}}{m_B + m_{K^*}}$ & $V^{B \to {K^*}}$ \\
         && $\bar{s} \gamma^\mu \gamma_5 h_v$  & $g^{\mu \nu}$ & $-i m_{K^*} f_B (m_B + m_{K^*})$ & $A_1^{B \to {K^*}}$ \\
         && $\bar{s} \gamma^\mu \gamma_5 h_v$  & $k^\mu q^\nu$ & $\frac{2 i m_{K^*} f_{K^*}}{m_B + m_{K^*}}$ & $A_2^{B \to {K^*}}$ \\
         && $\bar{s} \sigma^{\mu\{q\}} h_v$ &  $\varepsilon^{\mu \nu \{k q \}}$ & $2 i m_{K^*} f_{K^*}$ & $T_1^{B \to {K^*}}$ \\
         && $\bar{s} \sigma^{\mu\{q\}} \gamma_5 h_v$ & $q^\mu q^\nu$ & $2 m_{K^*} f_{K^*}$ & $T_{23B}^{B \to {K^*}}$ \\ \hline
         \multirow{ 5 }{*}{$\bar{B}^0 \to D^{*+}$}
         &\multirow{ 5 }{*}{$\bar{d} \gamma^\nu c$}
         & $\bar{c} \gamma^\mu h_v$ & $\varepsilon^{\mu \nu \{k q \}}$ & $\frac{2 m_{D^*} f_{D^*}}{m_B + m_{D^*}}$ & $V^{B \to {D^*}}$ \\
         && $\bar{c} \gamma^\mu \gamma_5 h_v$  & $g^{\mu \nu}$ & $-i m_{D^*} f_B (m_B + m_{D^*})$ & $A_1^{B \to {D^*}}$ \\
         && $\bar{c} \gamma^\mu \gamma_5 h_v$  & $k^\mu q^\nu$ & $\frac{2 i m_{D^*} f_{D^*}}{m_B + m_{D^*}}$ & $A_2^{B \to {D^*}}$ \\
         && $\bar{c} \sigma^{\mu\{q\}} h_v$ &  $\varepsilon^{\mu \nu \{k q \}}$ & $2 i m_{D^*} f_{D^*}$ & $T_1^{B \to {D^*}}$ \\
         && $\bar{c} \sigma^{\mu\{q\}} \gamma_5 h_v$ & $q^\mu q^\nu$ & $2 m_{D^*} f_{D^*}$ & $T_{23B}^{B \to {D^*}}$ \\
    \hline
    \end{tabular}
    \caption{Correspondence of form factors, processes, currents, Lorentz structures and $Y_F$ factors. We use the abbreviations $\sigma^{\mu \{ q \}} \equiv \sigma^{\mu \lambda} q_\lambda$ and $\varepsilon^{\mu \nu \{k q \}} \equiv \varepsilon^{\mu \nu \rho \sigma} k_\rho q_\sigma$  .}
    \label{tab:currents}
\end{table}

\subsection{Light-cone OPE of the correlation function} \label{sec:perturbation}

The next step to establish the LCSR is to compute the $B$ to vacuum correlation function \eqref{Correlation_function} in terms of $B$-meson LCDAs using a light-cone  operator product expansion (OPE)~\cite{Khodjamirian:1997lay}. To do so, we work in HQET in which the $b$ quark and $\bar{B}$ meson are substituted by the customary fields $h_v$ and $\ket{\bar{B}_v}$ defined as:
\begin{equation}
    h_v(x) \equiv e^{i m_b (v \cdot x)} \frac{1 + \slashed{v}}{2}b(x)\,, \quad \ket{\bar{B}_v} \equiv \frac{1}{\sqrt{m_B}} \ket{\bar{B}}\,,
\end{equation}
where $v = p_B/m_B$. We name the valence quarks of the final meson $q_1$, the interpolating quark emitted by the decay of the $b$ quark, and $q_2$, the spectator quark. Assuming that the external momenta are chosen such that $x^2 \sim 0$ and that the interpolating quark is highly virtual, the 2- and 3-particle contributions to the correlation function take the form \cite{Gubernari:2018wyi}:

\begin{equation}\label{2p}
    \Pi^{\mu \nu} \vert_{2p} =  \int d^4x \int \frac{d^4p'}{(2\pi)^4} e^{i(k-p').x} \left[ \Gamma_2^\nu \frac{\slashed{p'}+m_1}{m_1^2-p'^2}\Gamma_1^\mu \right]_{\alpha \beta}\bra{0}\bar{q_2}^\alpha(x) h_v^\beta(0) \ket{\bar{B}_v}\,, 
\end{equation}
\begin{align}\label{3p}
    \Pi^{\mu \nu} \vert_{3p} =  \int d^4x \int \frac{d^4p'}{(2\pi)^4} \int_0^1 du \; e^{i(k-p').x} &\left[ \Gamma_2^\nu \frac{(1-u)(\slashed{p'}+m_1)\sigma^{\lambda \rho}+u\sigma^{\lambda \rho}(\slashed{p'}+m_1)}{2(m_1^2-p'^2)^2}\Gamma_1^\mu \right]_{\alpha \beta} \nonumber\\
    &\times \bra{0}\bar{q_2}^\alpha(x) G_{\lambda \rho}(ux) h_v^\beta(0) \ket{\bar{B}_v}\,,  
\end{align}
where $G_{\lambda \rho}(ux)=g_s(\lambda^a/2)G^a_{\lambda \rho}(ux)$ is the gluon field strength tensor evaluated at $ux$, a fraction of the distance $x$. $\Gamma_{1,2}$ are defined by $J^\mu_{\text{weak}} \equiv \bar{q_1}(0)\Gamma^\mu_1 h_v(0)$ and $J^\nu_{\text{int}} \equiv \bar{q_2}(x)\Gamma^\nu_2 q_1(x)$ (see Table \ref{tab:currents}) and $m_1$ is the mass of the interpolating quark. We refer to Appendix \ref{appendix:B_LCDAs} for the parameterisation of the non-local $B$-to-vacuum matrix elements in Eqs.~\eqref{2p} and \eqref{3p} in terms of $B$-meson light-cone distribution amplitudes, following \cite{Grozin:1996pq, Braun:2017liq}.
For the 2-particle contribution, we work up to order $x^2$, corresponding to twists $2-5$. For the 3-particle one, we work at leading order in the light-cone OPE, including only twists $3$ and $4$. In this work, we derive explicitly the full expression for the twist-5 2-particle LCDA $g_-$ in Appendix \ref{appendix:B_LCDAs}. 

As in Eq.~\eqref{eq:Pi_p0}, by identification of Lorentz structures one can extract a scalar relation for each form factor:
\begin{equation}\label{eq:PiFPert}
     \Pi_F^{LCOPE}
    =  \int_0^{+\infty} d\sigma \sum_{n=1}^{+\infty}\frac{I_n^{F}(\sigma)}{(s(\sigma)-k^2)^n}\,, \\
\end{equation}
where $s$ is defined as $s(\sigma)=\sigma m_B^2+\frac{m_1^2-\sigma q^2}{1-\sigma}$ and $I_n^{F}(\sigma) = \sum\limits_{(\text{np})} I_n^{F (\text{np})}(\sigma)$ is a sum over the $n$-particle contributions. The 2-particle contributions are:
\begin{equation}
    I_n^{F (\text{2p})}(\sigma,q^2)
     = \frac{f_B m_B}{(1 - \sigma)^n} \sum_{\psi_\text{2p}}  C^{F, \psi_\text{2p}}_n(\sigma,q^2)\, \psi_\text{2p} (\sigma m_B)\,,
    \label{eq:CoeffFuncs2pt}\\ 
\end{equation}
where the summation goes over the 2-particle LCDAs defined in \eqref{2p-DAs} and $\sigma \equiv \omega/m_B$. The 3-particle contributions read:
\begin{align}
    I_n^{F(\text{3p})}(\sigma,q^2)
    = \frac{f_B m_B}{(1 - \sigma)^n}\int\displaylimits_{0}^{\sigma m_B}\text{d}\omega_1 \int \displaylimits_{\sigma m_B-\omega_1}^{\infty}\frac{\text{d}\omega_2}{\omega_2}
    \sum_{\psi_\text{3p}}  C^{F, \psi_\text{3p}}_n(\sigma,u,q^2)\, \psi_\text{3p} (\omega_1, \omega_2) \Bigg|_{u=(\sigma m_B-\omega_1)/\omega_2}\,,
\label{eq:CoeffFuncs3pt}
\end{align}
where the summation goes over the 3-particle LCDAs defined in \eqref{3p-DAs} and $\sigma \equiv (\omega_1+u\omega_2)/m_B$. We find that we are in agreement with \cite{Gubernari:2018wyi} regarding the perturbative coefficients $C_n^{F, \psi}$. 

We now estimate the condition on the kinematic variables of our problem such that the LCOPE is done in a perturbative regime. The LCOPE of the non-local matrix element is only valid near the light cone. Hence, the kinematical regime has to be such that the dominant contributions to \eqref{Correlation_function} arise from light-like distances, while respecting the conditions of QCD perturbativity given in Eqs.~(\ref{eq:perturbativity1}) and (\ref{eq:perturbativity2}) introduced below. To determine this regime, we note that the integral \eqref{Correlation_function} is dominated by the region where $k\cdot x \lesssim \mathcal{O}(1)$. We choose the frame of reference such that
\begin{equation}
    k\cdot x = k_0 x_0 - k_3 x_3\,.
\end{equation}
For $k^2 < 0$ we can write:
\begin{equation}
    k_0 = \frac{m_B^2 + k^2 - q^2}{2 m_B} , \hspace{1cm} \; k_3 = \sqrt{k_0^2 - k^2} , \hspace{1cm} \; x_3 = \frac{ k_0 x_0 - k \cdot x}{k_3}\,,
\end{equation}
which yields:
\begin{equation} \label{eq:x2}
    x^2  = - \frac{4 m_B \Bigl[ x_0^2 \, k^2 m_B - x_0 (k \cdot x)(k^2 + m_B^2 - q^2) + m_B(k \cdot x)^2 \Bigr] }{k^4 + (m_B^2 -q^2)^2 - 2 k^2 (m_B^2 + q^2)}\,.
\end{equation}
Using the expression of the LCDAs, we find that contributions to the correlation function at large $x_0$, typically a few $1/\Lambda_{QCD}$, are suppressed. Thus, keeping in mind that $x_0 \leq \mathcal{O}(1/\Lambda_{QCD})$ and $k\cdot x \lesssim \mathcal{O}(1)$, the dominant contributions to~\eqref{Correlation_function} arise on the light cone for any finite $q^2$ and a sufficiently large negative $k^2$. 
It is important to stress that this result is not practically useful for estimating the quality of the LCOPE. In order for the LCSR to have a predictive power, we need to suppress the contribution to the total correlation function from the \textit{a priori} unknown integral over the spectral density (second term in Eq.~\eqref{eq:Pi_p0}). While taking a large $k^2$ improves the LCOPE, it also enhances the relative size of this term and thus deteriorates the sum rule. The customary solution to this issue is to differentiate the total correlation function with respect to $k^2$ in order to suppress the contribution from the spectral density integral. However, such a differentiation enhances the relative size of the higher-order terms in the $x^2$ expansion, which breaks the LCOPE. Schematically, this can be seen in the relation $x^{2n} \propto 1/k^{2n}$ valid at large $k^2$. These two opposing effects are well-known and lead to the necessity of a compromise between the number of differentiation $p$ and the value of $k^2$. With these considerations we conclude that one cannot estimate the quality of the LCOPE purely from the kinematic variables $q^2$ and $k^2$. To estimate the quality of the LCOPE it is customary to simply make sure that the higher-twist contributions remain small relative to the leading twist contributions (see Section \ref{sec:estimation} for numerical details).  Generally, both the differentiation and the increase of $-k^2$ are performed simultaneously in the so-called Borel transformation, keeping $-k^2/p \equiv M^2$ constant and sending $-k^2,p\to \infty$. In this limit, the compromise is in the choice of the value of the Borel parameter $M^2$. As we show in the rest of this paper, applying the Borel transformation is not necessary and we prefer to keep $k^2$ and $p$ finite in order to track the dependence of our results on these parameters. 

We now turn to the conditions of perturbativity of QCD. We assumed that the interpolating quark $q_1$ is highly virtual and wrote the expression at leading order in QCD. Let us define the momentum exchange in HQET $\Tilde{q}=q - m_b v$. The condition for the interpolating quark to be highly virtual is
\cite{Khodjamirian:2020btr}:
\begin{equation} \label{eq:perturbativity1}
    \vert k^2 \vert , \vert \tilde{q}^2 \vert \gg \Lambda_{QCD}^2\,.
\end{equation}
The off-shellness of $q_1$ is not the only condition to ensure that the radiative corrections to $\Pi^{LCOPE}_F$ are small. Hard QCD effects arise from the interaction of the partons within the $B$ meson with themselves and with the interpolating quark. Part of these effects is absorbed by the LCDAs, whose scale dependence is known at NLO \cite{Braun:2017liq}. The full NLO calculation of $\Pi^{LCOPE}_F$ would, in principle, provide a robust quantification of the QCD perturbativity. However, it is not yet available for $B$-meson LCSRs in HQET. Hence we impose the following condition: the average virtuality in the correlator must remain well above the QCD scale: 
\begin{equation}\label{eq:perturbativity2}
    \vert \langle s \rangle - m_1^2 \vert \gg \Lambda_{QCD}^2\,.
\end{equation}
This is similar to the approach of \cite{Khodjamirian:2006st} which uses the Borel parameter $M^2$ as the QCD scale, as $M^2$ regulates the characteristic value of $s$ ($s \lesssim M^2$).
We define a way to estimate an average $\langle s \rangle$ in Section \ref{sec:estimation}. Obviously, the definition of an average $\langle s \rangle$ is somewhat arbitrary and a full NLO calculation would be needed to properly evaluate the range of applicability of perturbative QCD for this sum rule. 

\subsection{Quark-hadron duality} \label{sec:qhd}

Using \textit{global} quark-hadron duality, we can now equate both scalar expressions $\Pi_F$ \eqref{eq:Pi_p0} and $\Pi_F^{LCOPE}$ \eqref{eq:PiFPert}. To extract the hadronic form factor $F$, one still needs to estimate the integral over the density of the excited and continuum states which is achieved by using \textit{semiglobal} quark-hadron duality. The semiglobal quark-hadron duality postulates that for a sufficiently large negative $k^2$,
\begin{equation}\label{QH duality}
    \int_{s_{cont}}^{+ \infty}ds \frac{\rho_F(s)}{s-k^2} \approx \frac{1}{\pi}\int_{s_0}^{+ \infty}ds \frac{\text{Im}\Pi_F^{LCOPE}(s)}{s-k^2}\,,
\end{equation}
where $s_0$, the quark-hadron duality threshold, is an effective parameter \cite{Colangelo:2000dp}. The effective threshold is expected to be in the vicinity of the first excited or continuum state \cite{Colangelo:2000dp, Ball:2004rg, Bharucha:2015bzk}. Its estimation usually relies on the requirement that the value of the form factor is independent of $k^2$ (or, equivalently, the Borel parameter) which takes the form of a so-called \textit{daughter sum rule} (DSR) \cite{Ball:2004rg, Ball:2004ye, Bharucha:2015bzk, Gubernari:2018wyi}. In this work we define the daughter sum rule as the relation
\begin{equation} \label{eq:DSR}
    \frac{\partial}{\partial k^2} F(q^2) \propto \frac{\partial}{\partial k^2} \left((m_M^2 - k^2) \int_{0}^{s_0} ds \frac{\text{Im}\Pi_F^{LCOPE}(s)}{s-k^2} \right) \equiv 0\,,
\end{equation}
and equivalent relations obtained after differentiating the sum rule or applying the Borel transformation, which we solve to determine $s_0$. Alternatively, one may adopt thresholds from QCD sum rules \cite{Khodjamirian:2006st,Wang:2015vgv,Gubernari:2018wyi}. 

\subsection{Convergence of the LCOPE of the correlation function } \label{sec:pole}

The correlation function $\Pi^{LCOPE}_F$ \eqref{eq:Pi_p0} has a pole in $s(\sigma) = k^2$, which corresponds to the interpolating quark $q_1$ going on-shell. In fact, the quarks are confined in hadrons and cannot go on-shell \cite{Khodjamirian:2020btr}. We thus expect this to be a mathematical artifact. However, it needs to be discussed since we find that this pole can lead to the divergence of $\Pi^{LCOPE}_F$ in the scenario we detail below. For $k^2 < m_1^2$ the pole arises at:
\begin{equation}
    \sigma_{\text{pole}} = \frac{m_B^2 + k^2 - q^2 + \sqrt{(k^2 + m_B^2 - q^2)^2 + 4m_B^2(m_1^2 - k^2)}}{2 m_B^2}\,,
\end{equation}
which is close to unity for large $-k^2$. The LCDAs provided in Ref.~\cite{Braun:2017liq} come in three different models: exponential, local duality A and local duality B, which are asymptotically identical in the limit $\sigma \to 0$. For $q^2 = 0$, the local duality A and B models have cutoffs such that $I_n^F(\sigma > \sigma_0) = 0$ with $s(\sigma_0) > 0$ and $\sigma_0 < \sigma_{\text{pole}}$. Hence, for $k^2 < 0$ the pole is not reached in the integral \eqref{eq:PiFPert} and $\Pi^{LCOPE}_F$ converges. The exponential model on the contrary is defined with $\sigma \in \left[ 0, + \infty \right[ $. Hence, for $q^2 < m_1^2$, $s(\sigma)$ spans $\left[ m_1^2, + \infty \right[$ and $\left]-\infty, + \infty \right[$. In this case $\Pi^{LCOPE}_F$ diverges for any $k^2 \in {\rm I\!R}$. This prevents the use of the Cauchy integral formula on $\Pi^{LCOPE}_F$, which is needed to apply the QHD approximation. 
In order to circumvent this issue, one can perform a UV cutoff in the integral \eqref{eq:PiFPert} at $\sigma_{\text{max}} < \sigma_{\text{pole}}$. 
We can test this idea numerically, choosing $q^2 < m_1^2$ so that $\sigma_{\text{pole}} > 1$. We find that our results are independent of the choice of the cutoff for $\sigma_{\text{max}} \in \left[  0.75, 1 \right]$ which is due to the fact that LCDAs are mostly supported on the interval $\sigma \in \left[0,0.4\right]$. We also find that our numerical results employing the exponential model match the other two models within a few percent. This validates our strategy to regularise $\Pi_F^{LCOPE}$ in the exponential model.
$q^2 > m_1^2$ is more problematic as $s(\sigma)$ tends to $- \infty$ before the pole. In this case, the Cauchy integral theorem can only be applied for $k^2 > \text{max}(s(\sigma))$ which is incompatible with the sum rule method. A UV cutoff can regularise the integral and prevent $s$ from being negative, however, our results become more sensitive to said cutoff. We advocate for using $q^2 < m_1^2$ for LCSRs with $B$-meson LCDAs. In the rest of this work we set $q^2 = 0$.  

\subsection{Limitations of $B$-meson light-cone sum rules}
\label{sec:limitations}

$B$-meson LCSRs are a valuable tool for estimating form factors for final mesons whose distribution amplitudes are poorly known. They also conveniently provide correlated SM predictions between different $B$ decay channels. However, their accuracy is hindered by several factors that are often overlooked in the literature. We detail these factors in this section.

\textbf{Quark-hadron duality:} An essential step of establishing a sum rule is to approximate the hadronic correlation function by its partonic counterpart. This process, known as global quark-hadron duality, introduces a systematic and unknown error to the procedure. In addition to the global QHD, the semiglobal QHD presented in Eq. \eqref{QH duality} is an approximation which also introduces a theoretical uncertainty in the prediction of the form factors. We claim that this approximation can yield particularly large errors specifically for $B$-meson LCSR. The arguments for this claim are presented in the points below. In \ref{sec:method} we present a method to avoid the semiglobal QHD.

\textbf{LCDA models:} 
The functional dependence of the LCDAs on the internal parton momentum fraction $\omega$ is only known asymptotically in the limit $\omega \to 0$ \cite{Braun:1989iv}. Their behaviour at large momenta are phenomenologically constrained but can only be approximated by models (see Appendix \ref{appendix:B_LCDAs}). Using the three models considered in Ref.~\cite{Braun:2017liq}, we find that the prediction of form factors can be impacted by as much as $10 \%$ by the choice of the model. Recently, the authors of Refs.~\cite{Feldmann:2022uok,Feldmann:2023aml} proposed a systematic parameterisation of the leading $B$-meson light-cone distribution amplitude to improve on this issue.  
In addition, it is well-known that the higher twist contributions in the LCOPE are plagued by end-point divergences which can break the twist hierarchy (see discussion in \cite{Braun:2017liq}). 
Thus, our current understanding of $B$-meson LCDAs at high $\omega$ is quite approximate. Importantly, the semiglobal QHD approximation depends on this range to estimate the truncated spectral density integral. It is well-known that this dependence is suppressed by a small Borel parameter, and the upper end of the Borel window is chosen with this consideration in mind. In this work, we reevaluate the suppression of the contribution from the spectral density integral estimated using the semiglobal QHD \eqref{QH duality} in the Borel window used in the literature and advocate for smaller values of $M^2$ than previously considered in the literature (see Section \ref{sec:QHDvsUS}).

\textbf{LCDA parameters:} 
The LCDAs are expressed in terms of three input parameters: the inverse momentum $\lambda_B^{-1}$ and the parameters $\lambda_E^2$ and $\lambda_H^2$, which are non-perturbative quantities estimated using QCD sum rules. Because of our limited knowledge of the subleading quark condensates entering the QCD sum rule, the determination of these parameters comes with a large uncertainty.
The most critical parameter in $B$-LCSRs is $\lambda_B^{-1}$. It is scale dependent, and for conciseness, all values in this paragraph are taken at the scale $\mu = 1$ GeV. Estimates in the literature vary. For instance, Ref.~\cite{Braun:2003wx} reports $\lambda_B^{-1} = 2.15 \pm 0.5$ GeV$^{-1}$, similar to the $\lambda_B^{-1} = 2.2 \pm 0.6$ GeV$^{-1}$ used in Ref.~\cite{Gubernari:2018wyi}. More recent estimates from Ref.~\cite{Wang:2015vgv} present model-dependent results, with central values ranging from $2.57$ to $3.30$ GeV$^{-1}$. In this work, we use the latest estimation from \cite{Khodjamirian:2020hob}, which provides two estimates based on the model for quark-antiquark vacuum fluctuations for the strange quark condensate in the QCD sum rule: $2.17 \pm 0.24$ GeV$^{-1}$ and $3.05 \pm 0.56$ GeV$^{-1}$. Each prediction comes with a normally distributed error. The authors combine these results into a normally distributed $\lambda_B$. To maintain a normally distributed $\lambda_B^{-1}$, we perform our own combination. Our combination of the estimates from Ref.~\cite{Khodjamirian:2020hob} reads:
\begin{equation}
    \lambda_B^{-1} = 2.72 \pm 0.66 \, \textrm{GeV}^{-1}\,.
\end{equation}

In Tables \ref{tab:BtoVQHD} and \ref{tab:BtoPQHD}, we replicate the calculations from \cite{Gubernari:2018wyi}, comparing the predicted form factors using $\lambda_B^{-1} = 2.2 \pm 0.6$ GeV$^{-1}$ (GKvD \cite{Gubernari:2018wyi}, (i)) and $\lambda_B^{-1} = 2.72 \pm 0.66$ GeV$^{-1}$ ((ii), (iii) and (iv)). We observe a significant increase in all form factors when using the latter value, with up to a $50\%$ rise for $A_2$. This variation can have substantial phenomenological implications, indicating the need to reassess the determination of $\lambda_B^{-1}$. For example, in the case of $B \to K$ form factors, updating $\lambda_B^{-1}$ increases the form factors by $10\%$ and makes the predictions very close to other values found in the literature \cite{Parrott:2022rgu}. 

The parameters $\lambda_E^2$ and $\lambda_H^2$ are also poorly known. For definiteness, we use the estimation from \cite{Nishikawa:2011qk}. However, a more recent prediction from \cite{Rahimi:2020zzo}, using a different sum rule, finds a discrepant result. Using these newer values results in a $0-10\%$ variation in the prediction of the form factors.

\textbf{Range of the Borel parameter:} In the literature two ranges for the Borel parameter are commonly used for $B$-LCSR for $B$ decays to light mesons. The range $M^2 \in \left[0.5, 1.5\right]$ GeV$^2$ \cite{Khodjamirian:2006st,Gubernari:2018wyi} is taken from two-point QCD sum rules \cite{Shifman:1978bx,Colangelo:2000dp} and LCSRs with the pion distribution amplitude \cite{Braun:1994ij,Khodjamirian:1997tk,Braun:1999uj}. 
Alternatively, the authors of Refs.~\cite{Wang:2015vgv, Lu:2018cfc, Cui:2022zwm} and \cite{Descotes-Genon:2019bud, Descotes-Genon:2023ukb} use $M^2 \in [1, 1.5]$ GeV$^2$, based on several criteria: ensuring the spectral density integral remains small, minimising the form factors' dependence on the Borel parameter, maintaining a threshold close to the one estimated from two-point QCD sum rules, and suppressing higher twist contributions.
The choice of the Borel window significantly influences the magnitude of the predicted form factor. As shown in Table \ref{tab:QHD}, there is a $20\%$ variation between the extremes of the window $\left[0.5, 1.5\right]$ GeV$^2$. While one expects the form factors to be independent of the $M^2$, it is clear that the choice of the Borel window is crucial for the final result of the sum rule. 

\textbf{Determination of the effective threshold $\boldsymbol{s_0}$:} 
In Ref.~\cite{Gubernari:2018wyi}, it has been found that for ${B\to\pi,\rho,K}$ the determination of the effective threshold following a strategy similar to the daughter sum rule method described in Section~\ref{sec:qhd} fails. For this reason, the author relied on threshold obtained from QCD sum rules.  We are able to reproduce this observation for $B \to \pi, K$, for which the DSR yields effective thresholds much lower than the range expected from two-point QCD sum rules, and yields very small predictions for the form factors (see Table~\ref{tab:BtoPQHD} (iii)). For $B \to \rho$, we also find a smaller effective threshold, yet still in the ballpark of the one hinted at by QCD sum rules. By combining the re-evaluation of the effective threshold and the updated value of $\lambda_B^{-1}$, we reproduce (coincidentally) exactly the prediction of Ref.~\cite{Gubernari:2018wyi} (see Table~\ref{tab:BtoVQHD}).

For $B \to K^*$, we find that the predictions of the form factors are crucially dependent on the strategy adopted to determine the effective threshold $s_0$. As a demonstration, we reproduce exactly the calculation of Ref.~\cite{Gubernari:2018wyi} and use the same input parameters, changing only the way we determine the effective threshold (see Table~\ref{tab:BtoVQHD}). In our implementation, we solve Eq.~\eqref{eq:DSR} for $s_0$ for each set of input parameters sampled from our prior. We find results compatible with \cite{Gubernari:2018wyi}, however, there is up to $8 \%$ of variation in the final results using exactly the same parameter input and calculations. More importantly, our procedure more than doubles the size of the theoretical error on the predicted form factors. 

\begin{table}[!t]
    \centering
    \begin{tabular}{|c|c|c|c|c|c|c|}
        \hline
        Form & \multicolumn{3}{c|}{$B \to K^*$} & \multicolumn{3}{c|}{$B \to \rho$} \\ \cline{2-7}
         factor & GKvD \cite{Gubernari:2018wyi} & (i) & (ii) & GKvD \cite{Gubernari:2018wyi} & (i) & (ii) \\ \hhline{-|---|---}
        $V$  & $0.33 \pm 0.11$ & $0.31_{-0.15}^{+0.19}$& $0.48_{-0.20}^{+0.24}$ & $0.27 \pm 0.14$ & $0.16_{-0.09}^{+0.12}$ & $0.27_{-0.13}^{+0.16}$\\
        $A_1$ & $0.26 \pm 0.08$ & $0.25_{-0.12}^{+0.14}$ & $0.36_{-0.15}^{+0.18}$ & $0.22 \pm 0.10$ & $0.14_{-0.07}^{+0.09}$ & $0.21_{-0.10}^{+0.11}$ \\
        $A_2$ & $0.24 \pm 0.09$ & $0.22_{-0.12}^{+0.16}$ & $0.36_{-0.17}^{+0.20}$ & $0.19\pm 0.11$ & $0.11_{-0.07}^{+0.10}$ & $0.20_{-0.10}^{+0.14}$ \\
        $T_1$& $0.29 \pm 0.10$ & $0.27_{-0.13}^{+0.17}$ & $0.41_{-0.17}^{+0.20}$ & $0.24 \pm 0.12$ & $0.15_{-0.08}^{+0.10}$ & $0.24_{-0.11}^{+0.13}$\\
        $T_{23}$ & $0.58 \pm 0.13$ & $0.58_{-0.20}^{+0.19}$ & $0.73_{-0.21}^{+0.16}$ & $0.56 \pm 0.15$ & $0.43_{-0.15}^{+0.16}$ & $0.56_{-0.16}^{+0.16}$\\ \hline
        $s_0$ (GeV$^2$)& $\left[ 1.4, 1.7\right]$ & $1.53_{-0.09}^{+0.35}$ & $ 1.54_{-0.10}^{+0.34}$ & $1.6 \pm 0.032$ & $1.03_{-0.04}^{+0.08}$ & $1.05_{-0.04}^{+0.09}$\\ \hline
    \end{tabular}
    \caption{Prediction of $B \to \rho, K^*$ form factors at $q^2 = 0$ following the calculation of \cite{Gubernari:2018wyi}. We include the original results from \cite{Gubernari:2018wyi} and our results using a different method for the determination of the effective threshold, obtained with $\lambda_B^{-1} = 2.2 \pm 0.6$ GeV$^{-1}$ (i) and ${\lambda_B^{-1} = 2.72 \pm 0.66}$~GeV$^{-1}$ (ii).}
    \label{tab:BtoVQHD}
\end{table}

\begin{table}[!t]
    \centering
    \begin{tabular}{|c|c|c|c|c|c|c|}
        \hline
        Form & \multicolumn{3}{c|}{$B \to \pi$} & \multicolumn{3}{c|}{$B \to K$} \\ \cline{2-7}
         factor & GKvD \cite{Gubernari:2018wyi} & (iii) & (iv) & GKvD \cite{Gubernari:2018wyi} & (iii) & (iv) \\ \hhline{-|---|---}
         $f_+$ & $0.21(7)$ & $0.023(7)$ & $0.26_{-0.08}^{+0.08}$ & $0.27(8)$ & $0.24(7)$  & $0.34_{-0.09}^{+0.09}$ \\
         $f_T$ & $0.19(7)$ & $0.024(7)$ & $0.24_{-0.06}^{+0.06}$ & $0.25(7)$ & $0.24(7)$ & $0.31_{-0.08}^{+0.06}$ \\ \hline
         $s_0$ (GeV$^2$) & $0.7 \pm 0.014$ & $0.0393(1)$ & $0.7 \pm 0.014$ & $1.05 \pm 0.021$ & $0.54_{-0.02}^{+0.03}$ &  $1.05 \pm 0.021$ \\ \hline
    \end{tabular}
    \caption{Prediction of $B \to \pi, K$ form factors at $q^2 = 0$ following the calculation of \cite{Gubernari:2018wyi}. We include the original results from \cite{Gubernari:2018wyi} and our results obtained using $\lambda_B^{-1} = 2.72 \pm 0.66$ GeV$^{-1}$ and $s_0$ obtained from a daughter sum rule (iii) and using the same threshold $s_0$ as in \cite{Gubernari:2018wyi} (iv).}
    \label{tab:BtoPQHD}
\end{table}

\vspace{1cm}
The list of these issues demonstrates that $B$-meson LCSR needs to be improved on many aspects. In this article, we present a method to tackle three of these issues, namely, the determination of the effective threshold, the choice of the Borel parameter and the error from semiglobal QHD. We explore the potential of LCSRs in a regime where the semiglobal quark-hadron duality is not needed therefore removing the associated systematic uncertainty. We show that this approach is viable, at the price of a higher dependence on the radiative corrections and higher twists in the correlation function.

\section{Light-cone sum rules without semiglobal quark-hadron duality}
\subsection{Presentation of the method}\label{sec:method}

\begin{figure}[H]
    \centering
    \includegraphics[width=\textwidth]{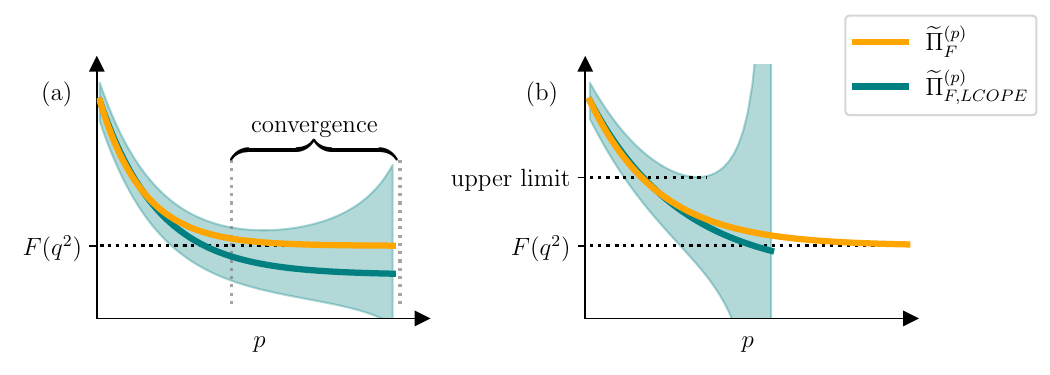}
    \caption{Schematic illustration of the strategy.}
    \label{fig:method_illustration}
\end{figure}

Our objective is to remove the contribution from the spectral density integral \eqref{QH duality} in the expression of the correlation function $\Pi_F$ without employing the semiglobal QHD approximation. Following the method of power moments in QCD sum rules \cite{Shifman:1978bx} we take the $p$-th derivative of Eq.~\eqref{eq:Pi_p0} with respect to $k^2$ \cite{Khodjamirian:2020btr}:

\begin{equation}
    \Pi_F^{(p)}(q^2,k^2) \equiv  \left( \frac{\partial}{\partial k^2} \right)^p \Pi_F(q^2,k^2) = \, p! \left( Y_F \frac{F(q^2)}{(m_M^2 - k^2)^{p+1}} + \int_{s_{cont}}^\infty \frac{\rho_F(s)}{(s-k^2)^{p+1}} \right)\,.
\end{equation}
The form factor $F(q^2)$ can then be rewritten as:
\begin{equation} 
    F(q^2) = \frac{(m_M^2 - k^2)^{p+1}}{p! \, Y_F} \Pi_F^{(p)}(q^2,k^2) - \int_{s_{cont}}^\infty \frac{\rho_F(s)}{Y_F} \left( \frac{m_M^2 - k^2}{s-k^2} \right)^{p+1}\,.
\end{equation}
Since $m_M^2 < s_{cont}$ and $k^2<0$,
\begin{equation}
    \int_{s_{cont}}^\infty \frac{\rho_F(s)}{Y_F} \left( \frac{m_M^2 - k^2}{s-k^2} \right)^{p+1} \xrightarrow[p \to \infty]{} 0\,,
\end{equation}
hence, the form factors can be expressed as:
\begin{equation}\label{eq:lim_FF}
    F(q^2) = \lim_{p\to\infty}  \frac{(m_M^2 - k^2)^{p+1}}{p! \, Y_F} \Pi_F^{(p)}(q^2,k^2)\,.
\end{equation}
This expression is exact and does not rely on semiglobal QHD. A corollary expression can be derived with the same approach. The squared mass of the final meson as a function of the correlation function is:
\begin{equation}\label{eq:lim_m2}
        m_M^2 = \lim_{p\to\infty}  \left[ \frac{p!}{(p-\ell)!}  \frac{\Pi_F^{(p-\ell)}}{\Pi_F^{(p)}} \right]^{1 / \ell} + k^2, \quad p>1 , \; p> \ell \geq 1 \, .
\end{equation}
This latter expression originates from the same relation between the derivative of the correlation function and the meson mass $m_M$ used in daughter sum rules \cite{Khodjamirian:2020btr}. For brevity of notation, we define:
\begin{equation}
    \widetilde{\Pi}_F^{(p)}(q^2,k^2) \equiv \frac{(m_M^2 - k^2)^{p+1}}{p! \, Y_F} \Pi_F^{(p)}(q^2,k^2), \quad R_F(p,q^2,k^2) \equiv \int_{s_{cont}}^\infty \frac{\rho_F(s)}{Y_F} \left( \frac{m_M^2 - k^2}{s-k^2} \right)^{p+1}\,,
\end{equation}
and 
\begin{equation} \label{eq:def_mtilde}
    \widetilde{m}_M^2(p,\ell,k^2) \equiv \left[ \frac{p!}{(p-\ell)!}  \frac{\Pi_F^{(p-\ell)}}{\Pi_F^{(p)}} \right]^{1 / \ell} + k^2\,,
\end{equation}
such that:
\begin{equation}
        \widetilde{\Pi}_F^{(p)}(q^2,k^2) = F(q^2) + R_F(p) \,,
\end{equation}
and 
\begin{equation} 
    \widetilde{\Pi}^{(p)}_F(q^2,k^2) \xrightarrow[p \to \infty]{} F(q^2), \quad R_F(p,q^2,k^2) \xrightarrow[p \to \infty]{} 0, \quad \widetilde{m}_M^2(p,\ell,k^2) \xrightarrow[p \to \infty]{} m_M^2\,.
\end{equation}
The relations we have derived thus far in this section are exact, but of little use without the knowledge of the correlation function $\Pi_F$ to all orders in the LCOPE and in perturbation theory. In practice, we can only approximate the correlation function with an expression expanded on the light cone, as detailed in Section \ref{sec:perturbation}. The difference between the 'true' and the LCOPE expressions of the correlation function originates from the truncation of the multiple expansions performed to obtain $\Pi_{F,LCOPE}^{(p)}$. This truncation error grows as $p$ increases because both the higher order LCOPE and hard QCD corrections become large, as we discuss in the next section. We plot schematically this behaviour in Fig.\ref{fig:method_illustration}. Our strategy is the following: We quantify the $p$-dependent truncation error in Section \ref{sec:estimation} with a conservative order of magnitude estimate such that $\Pi_F^{(p)} \approx \Pi_{F,LCOPE}^{(p)}$ within uncertainties for all $p$. Necessarily, when $p$ becomes too large, the estimated uncertainty diverges, and the calculation loses its predictive power. We claim that if the truncation error is not too large at large $p$ useful information can be extracted from LCSRs without semiglobal QHD. 

In fact, when $p$ increases there are two possible outcomes:
\begin{enumerate}[label=(\alph*)]
    \item The spectral density integral $R_F$ becomes negligible compared to $F(q^2)$ and to the total error on $\widetilde{\Pi}_{F,LCOPE}^{(p)}$ before the latter diverges. We call this regime 'convergence of the sum rule' (see Fig.\ref{fig:method_illustration} (a)). In this case, we obtain a prediction of the form factor which does not require the knowledge of $R_F$. In order to know that we have reached this regime, there are two criteria we can use without relying on the semiglobal QHD approximation:
    \begin{itemize}
        \item If truncation errors are small enough and the error is dominated by the LCDAs inputs then $\widetilde{\Pi}_{F,LCOPE}^{(p)}$ and $\widetilde{m}_M^2$ become weakly dependent on $p$ and $k^2$ near the convergence.
        \item In Eq.~\eqref{eq:lim_m2} $m_M$ is not an input of the mass sum rule, it is a genuine prediction of the meson mass from first principles. Usually, this relation is used to set the effective threshold in semiglobal QHD \cite{Khodjamirian:2020btr}, instead we use it as a criterion of convergence. A Taylor expansion of Eq.~\eqref{eq:def_mtilde} yields:
        \begin{equation}
            \widetilde{m}_M^2 = m_M^2 + (m_M^2 - k^2) \left[ \frac{1}{\ell} \cdot \frac{R_F(p-\ell) - R_F(p)}{F(q^2)} + \mathcal{O}\left(\frac{R_F}{F}\right)^2\right] \,.
        \end{equation} 
        Hence, assuming that $R_F$ does not plateau before going to zero, the convergence of $\widetilde{m}_M^2 \to m_M^2$ is a proxy for the quality of the convergence of the sum rule.
    \end{itemize}
    In this regime, attempting to estimate a negligible $R_F$ with semiglobal QHD might introduce a systematic error that can be larger than the actual size of $R_F$. This approach addresses some of the points raised in Section \ref{sec:limitations} as it does not rely on the determination of the effective threshold. Furthermore, it yields a reduction of the dependence on the choice of the LCDA model in the large $p$ limit.
    \item The error in the correlation function diverges before reaching the regime described above (see Fig.\ref{fig:method_illustration} (b)). In this situation we can use the semilocal quark-hadron duality to determine the sign of $R_F$ for all considered $p$ using daughter sum rules to estimate the effective thresholds. From this information we can deduce an upper or lower limit on $F(q^2)$. In practice, we find $R_F > 0$ in most cases, and in the few cases where $R_F<0$ the corresponding lower limits are negative, hence we focus on upper limits only. In principle, these upper limits are robust results since we rely on semiglobal QHD solely to test the sign of $R_F$. While not as compelling as predictions, these upper limits on local form factors can be useful in certain contexts. For example, smaller SM predictions for $f_{+}^{B\to K}(q^2)$ would reduce the tension between the experimental measurements and the SM predictions of $\text{BR}(B^+ \to K^+ \mu \mu)$. The determination of upper limits with sum rules is not a new idea. It has already been considered in the case of QCD sum rule, where the spectral density $\rho(s)$ is positive definite. This consideration has been applied in the prediction of LCDA parameters, as documented in Ref.~\cite{Rahimi:2020zzo}, as well as in the determination of decay constants, as discussed in Refs.~\cite{Gelhausen:2013wia, Khodjamirian:2008xt}.  
\end{enumerate}

In the usual LCSR method using the Borel transformation, the relations \eqref{eq:lim_FF} and \eqref{eq:lim_m2} correspond to the limit $M^2 \to 0$. In this work, we avoid the Borel transformation for two reasons. First, the convergence criteria we introduced in this section include the independence of $\widetilde{\Pi}_F^{(p)}$ on both $k^2$ and $p$ independently. This is a more stringent condition than the independence on their ratio $M^2$ which is used in the traditional LCSR approach (see e.g. \cite{Colangelo:2000dp}). The second advantage of keeping an explicit dependence on $k^2$ and $p$ is that the convergence is sometimes better at low $-k^2$ (although marginally) than at larger $-k^2$ where we recover the result of the Borel transformation. This is illustrated in Table \ref{tab:resultsBtoP} for the case of the $B \to \pi$ form factors. 

A motivation for using the Borel transformation is that it removes subtraction terms that could be present in dispersion relations of sum rules. We note that this property is not exclusive to the Borel transformation since for a finite number $n$ of subtractions in a dispersion relation, it is sufficient to perform $n$ differentiations to remove the subtraction terms \cite{Khodjamirian:2020btr}. The sum rule studied in this work is not regulated by subtractions, hence we do not need to consider this issue.

We find numerically that the convergence of the Borel transformation when taking the limit $-k^2,p \to + \infty$ happens for $-k^2$ of the order of a few GeV when $-k^2/p \sim 1$ GeV$^2$. Since we set $-k^2 \geq 2$ GeV$^2$ (see discussion in Section \ref{sec:radiative_error}), the numerical analysis we perform yields results close to what one would obtain using the Borel transform. Thus, in the rest of this article we will be able to compare the ratio $-k^2/p$ and the Borel parameter $M^2$ used in other similar calculations.

In order for this approach to yield useful results, we need a conservative yet accurate estimation of the truncation error. We detail our estimation of the theoretical error coming from the multiple truncations in the expression of $\Pi_{F, \,LCOPE}^{(p)}$ in Section \ref{sec:estimation} and present our numerical results in Section~\ref{sec:numerics}. In this work, we apply this method to $B$-meson LCSRs but the same strategy can be applied to LCSRs with light-meson LCDAs as shown in Appendix \ref{appendix:lightmesonDAs}. The advantage of $B$-meson LCSRs over light-meson LCSRs is the large number of form factors accessible with a single calculation, which allows us to check our results against numerous other works.

\subsection{Truncation error in $\Pi_{F, \,LCOPE}^{(p)}$} \label{sec:estimation}
In this section, we estimate the uncertainties associated with the truncations of all expansions in the expression of $\Pi_{F, \,LCOPE}^{(p)}$. Our strategy is to write a parametric expression of the true correlation function in terms of $\Pi_{F, \,LCOPE}^{(p)}$ and unknown parameters whose prior we establish conservatively. Schematically we write $\Pi_{F}^{(p)} = \Pi_{F, \,LCOPE}^{(p)} + w \Delta(p)$, where $\Delta(p)$ is the estimated size of the correction and $w$ is a parameter of order one.
\subsubsection{Fock state expansion in \textit{\textbf{n}}-particle contributions}
We perform the calculation up to $3$-particle contributions. We define $w_{\textrm{n-part}}$ such that:
\begin{equation} \label{eq:nparterror}
\begin{split}
    \Pi_{F}^{(p)} = \sum_{\text{n} = 2}^{+\infty} \Pi_{F,LCOPE}^{(p)} \vert_{(\textrm{n-part})}& = \Pi_{F,LCOPE}^{(p)} \vert_{(\textrm{2-part})} +  \Pi_{F,LCOPE}^{(p)} \vert_{(\textrm{3-part})} + \delta_{(\textrm{part})}^{pert(p)} \\
    & =\Pi_{F,LCOPE}^{(p)} \vert_{(\textrm{2-part})} + \Pi_{F,LCOPE}^{(p)} \vert_{(\textrm{3-part})} (1 + w_{\textrm{n-part}}^{(p)})\,.
\end{split}
\end{equation}
Numerically, we check that $\Pi_{F,LCOPE}^{(p)} \vert_{(\textrm{3-part})}\ll \Pi_{F,LCOPE}^{(p)} \vert_{(\textrm{2-part})}$ for the values of $p$ taken in consideration, which is in line with the results of \cite{Gubernari:2018wyi}. We choose $w_{\textrm{n-part}}^{(p)}$ to be uniformly distributed in $\left[ -2,2\right]$, which we deem conservative since the truncated terms are suppressed in the $n$-particle expansion.

\subsubsection{Radiative corrections in $\alpha_s$} \label{sec:radiative_error}
In Section~\ref{sec:perturbation}, we discussed the kinematic conditions for the interpolating quark $q_1$ to be highly virtual and for the radiative QCD corrections to remain small, and found $\vert k^2 \vert, \vert \tilde{q}^2 \vert$, ${\langle s \rangle - m_1^2 \gg \Lambda_{QCD}^2}$. We define the average value of $s$ as:
\begin{equation} \label{eq:avgs}
    \langle s \rangle = \frac{ \int_0^{\sigma_{\textrm{max}}} d \sigma \vert I_{\textrm{tot}}^{(F,p)}(\sigma, k^ 2) \vert s(\sigma) }{\int_0^{\sigma_{\textrm{max}}} d \sigma \vert I_{\textrm{tot}}^{(F,p)}(\sigma, k^2)\vert}\,,
\end{equation}
where $I_{\textrm{tot}}^{(F,p)}(\sigma, k^ 2)$ is the integrand of the $p-$th derivative of the correlation function calculated with the LCOPE \eqref{eq:PiFPert}:
\begin{equation} \label{eq:PiFpert_p}
    \Pi_{F,LCOPE}^{(p)}(q^2,k^2) = \int_0^{\sigma_{\textrm{max}}} d \sigma \sum_{n = 1}^{\infty} \frac{(n+p-1)!}{(n-1)!} \frac{I^{(F)}_n(\sigma)}{(s(\sigma) - k^2))^{n+p}} \equiv \int_0^{\sigma_{\textrm{max}}} d \sigma I_{\textrm{tot}}^{(F,p)}(\sigma, k^2)\,.
\end{equation}
For $p=0$, $\langle s \rangle$ is of the order of a few GeV$^2$, and when $p \to \infty$, $\langle s \rangle \to m_1^2$. To estimate the size of the radiative correction, we define the characteristic scale
\begin{equation}
    \mu_{\textrm{QCD}} \equiv \text{min}(\sqrt{\langle s \rangle - m_1^2},\sqrt{\vert k^2 \vert}, \sqrt{\vert \tilde{q}^2 \vert})\,.
\end{equation}
For a given renormalisation scheme, the correlation function expanded to all orders in perturbative QCD can be written as 
\begin{equation}\label{eq:QCDerror}
    \Pi^{(p)}_{F} = \Pi^{(p)}_{F,LO} \left[ 1 + w_{\alpha_s}^{(p)}(\mu) \sum_{n=1} \left( \frac{\alpha_s(\mu)}{\pi} \right)^n \right]  = \Pi^{(p)}_{F,LO} \left[ 1 + w_{\alpha_s}^{red}{(p)}(\mu) \, \frac{\alpha_s(\mu)/\pi}{1 - \alpha_s(\mu)/\pi} \right]\,,
\end{equation}
where $w_{\alpha_s}(\mu \sim m_B) \sim 1$ and $w_{\alpha_s}(\mu \to \Lambda_{QCD}) = 0$. We choose to work in the $\overline{\textrm{MS}}$ renormalisation scheme and set $\mu = \mu_{QCD}$. Setting $w_{\alpha_s}$ to be of order $1$ and independent of the scale $\mu$ gives a crude yet conservative estimation of the QCD error when $p \to \infty$.
Radiative corrections at NLO have been calculated for $B$-meson LCDAs in SCET in Ref.~\cite{Cui:2022zwm}. In that work the Borel parameter is set to $M^2 = 1.25 \pm 0.25 \textrm{ GeV}^2$ and the authors find that the largest radiative corrections are of order $30 \%$. Recalling that $s \lesssim M^2$ and setting the scale $\mu = \sqrt{1.25 \textrm{ GeV}^2}$, our estimation of the missing radiative correction is $w_{\alpha_s} \times \alpha_s(\mu)/(\pi - \alpha_s(\mu)) = 0.25 w_{\alpha_s}$. Hence, we choose $\left[-1.5,1.5\right]$ as a conservative interval for $w_{\alpha_s}$. Our QCD error estimation does not account for possible large logarithms that may appear. However they have been included and resummed in Ref.~\cite{Cui:2022zwm}, and are part of the quoted $30\%$ error. In the mass sum rule \eqref{eq:lim_m2}, we make the conservative assumption that radiative corrections are uncorrelated between different derivatives. 

\subsubsection{LCOPE truncation error}

We now focus on assessing the error arising from truncating the LCOPE. Matching the twist expansion of the LCDAs with that of the LCOPE, in the case of 2-particle LCDAs, the leading twists (LT) $2$ and $3$ contribute at order $(x^2)^0$, while the next-to-leading twists (NLT) $4$ and $5$ contribute at order $(x^2)^1$ and so forth. For $n$-particle contributions, the twist counting starts at $n$. Schematically we have:
\begin{equation} \label{eq:LCOPEexpansion}
    \Pi_{F, LCOPE}(q^2,k^2) = \underbrace{\Pi_{2p}^{twist-2,3} + \Pi_{3p}^{twist-3,4}}_{\propto (x^2)^0 \text{: LT}}+ \underbrace{\Pi_{2p}^{twist-4,5} + \Pi_{3p}^{twist-5,6}}_{\propto x^2 \text{: NLT}} + \underbrace{\Pi_{2p}^{twist-6,7}+\Pi_{3p}^{twist-7,8}}_{\propto x^4 \text{: NNLT}} + ... \,.
\end{equation}
Additionally, the LCDAs are expanded within HQET. As well known, the power expansion in HQET is mismatched with the expansion in twists, hence the power counting in HQET is ill-defined at a given order in the twist expansion. However, we know that order $x^2$ in the LCOPE expansion corresponds to order $1/m_b + \mathcal{O}(1/m_b^2)$ in HQET, which can be seen e.g. from Eq.~\eqref{eq:x2}. We include 2-particle $B$-meson LCDAs up to twist-5, thus encompassing the entire $\mathcal{O}(1/m_b)$ order of HQET. For 3-particle LCDAs, we only consider contributions up to twist-3 and twist-4, which correspond to the leading order contribution in the LCOPE and leading power in HQET. At the leading order in QCD, the propagator of the interpolating quark $q_1$ is exact at all orders in HQET, hence the order $1/m_b$ is fully included at QCD leading order in the 2-particle contributions. We do not account for the effect of higher power corrections which we expect to be negligible compared to the other errors we take into account. Following the series expansion introduced in Eq. \eqref{eq:LCOPEexpansion}, we can write the correlator as: 
\begin{equation}
     \Pi_{F}^{(p)} = \sum_{t \geq 2} \Pi_{twist=t}^{(p)} =  \Pi_{\text{LT}}^{(p)} + \Pi_{\text{NLT}}^{(p)} + \Pi_{\text{NNLT}}^{(p)} + ... \,.
\end{equation}
We work under the assumption that all coefficients multiplying powers of $x^2$ in the light-cone OPE are of the same order of magnitude as the leading twist contribution:
\begin{equation}
    \frac{\Pi_{k\textrm{NLT}}^{(p)}}{\Pi_{\textrm{LT}}^{(p)}} \frac{1}{x^{2k}} = \mathcal{O}(1)\,.
\end{equation}
To be conservative, we assume that all the terms in the series interfere positively such that:
\begin{equation}
    \Big\vert \sum_{t \geq 6} \Pi_{twist=t}^{(p)} \Big\vert \approx \Big\vert \frac{\Pi_{\textrm{NLT}}^2}{\Pi_{\textrm{LT}} - \Pi_{\textrm{NLT}}} \Big\vert \leq  \frac{\Pi_{\textrm{NLT}}^2}{\vert \Pi_{\textrm{LT}} \vert - \vert \Pi_{\textrm{NLT}} \vert} \,.
\end{equation}
Following this approximation, we estimate the contribution of the missing twists by: 
\begin{equation} \label{eq:LCOPEerror}
    \begin{split}
   \Pi_{F}^{(p)} \equiv \Pi_{\text{LT}}^{(p)} + \Pi_{\text{NLT}}^{(p)} + w_{LCOPE} \times \frac{(\Pi_{\text{NLT}}^{(p)})^2}{\vert \Pi_{\text{LT}}^{(p)} \vert - \vert \Pi_{\text{NLT}}^{(p)} \vert }\,,
   \end{split}
\end{equation}
where $w_{LCOPE}$ is an unknown parameter of order unity. We take it to be uniformly distributed in the range $\left[-2,2\right]$. We also ensure that $\Pi_{\text{NLT}}^{(p)} / \Pi_{\text{LT}}^{(p)} \leq 30 \%$. Numerically, we find that the size of the uncertainty coming from the truncation of the LCOPE using this model is very small compared to the total uncertainty, which is dominated by parametric uncertainties and radiative corrections (this is also visible in  Fig.~\ref{fig:FFs_correlation} where the error coming from the above model is called "twist", and contributes a small fraction of the total error in $f_+^{B \to \pi,K}$ and $A_1^{B \to \rho}$). Doubling the size of the allowed interval for $w_{LCOPE}$ to $\left[-4, 4\right]$ yields an increase of the total relative uncertainty of about $1\%$. At large $p$ and small $\Pi_{\text{NLT}}^{(p)} / \Pi_{\text{LT}}^{(p)}$ one can show that $w_{LCOPE}$ is weakly dependent on $p$. We use that approximation of $p$-independent $w_{LCOPE}$ in the mass sum rule \eqref{eq:lim_m2}.
From the expression in Eq.~\eqref{eq:PiFpert_p}, one can infer how the quality of the LCOPE is regulated by $k^2$ and $p$. In Eq.~\eqref{eq:PiFpert_p}, the leading twist contributes to $I^{(F)}_{n=1,2}(\sigma)$ and next-to-leading twist contributes to $I^{(F)}_{n=2,3,4}(\sigma)$. Hence, the ratio of the two successive terms in \eqref{eq:PiFpert_p} roughly corresponds to the ratio of next-to-leading twist over leading twist contributions. This ratio goes like $(n+p)/n \times 1/(s(\sigma) - k^2)$, which in the limit of $n \ll p$ and $-k^2 \gg \langle s \rangle$ goes like $p/-k^2$. This shows qualitatively how $p/-k^2$ and the Borel parameter regulate the goodness of the LCOPE.

\subsubsection{Total error model} 
Combining Eqs.~\eqref{eq:nparterror}, \eqref{eq:QCDerror} and \eqref{eq:LCOPEerror}, we obtain the following expression for the exact correlation function:

\begin{equation}  \label{eq:expansion_error}
\begin{aligned}
    \Pi_{F}^{(p)} = \left( 1 +  w_{\alpha_s}^{(p)} \times \frac{\alpha_s(\mu_{\text{QCD}})/\pi}{1 - \alpha_s(\mu_{\text{QCD}})/\pi} \right) \times \Biggr[ \sum_{\textrm{twist} = \text{LT}, \text{NLT}} &\left[ (\Pi^{2p}_{LO})^{(p)} + (\Pi^{3p}_{LO})^{(p)} (1 + w_{n\textrm{-part}}^{(p)})) \right]
 \\
 & + w_{LCOPE} \times \frac{(\Pi_{LO, \text{NLT}}^{(p)})^2}{ |\Pi_{LO, \text{LT}}^{(p)}| - |\Pi_{LO, \text{NLT}}^{(p)}|} \Biggr]\,.  
\end{aligned}
\end{equation}

We treat $w_{n\text{-part}}^{(p)},w_{\alpha_s}^{(p)},w_{LCOPE}$ as nuisance parameters in a frequentist inference to account for the total error from the multiple truncations. Based on the discussion above, we assume that they are uniformly distributed in the following ranges:
\begin{equation} \label{eq:error_param}
    w_{n\text{-part}}^{(p)} \in \left[-2,2\right], \quad w_{\alpha_s}^{(p)} \in \left[-1.5, 1.5\right], \quad w_{LCOPE} \in \left[-2,2\right]\,.
\end{equation}
 In this simplified model, as $p$ increases, the estimated uncertainty on $\Pi_{F}^{(p)}$ becomes out of control because of the simultaneous decrease of $\mu_{QCD}$ toward zero and the increase of $\Pi_{\text{NLT}}^{(p)}/\Pi_{\text{LT}}^{(p)}$ toward unity. 

\section{Numerical Results} \label{sec:numerics}
\subsection{Input parameters} \label{sec:input}
We use the PDG values for the meson masses, while the remaining input parameters are listed in Table \ref{tab:input_param}. The running of $\alpha_s(\mu)$ in the $\overline{\textrm{MS}}$ renormalisation scheme is performed using the package \texttt{rundec} \cite{Herren:2017osy}. We work in a frequentist inference and sample our expression for $\Pi_{F}^{(p)}$ from Eq.~\eqref{eq:expansion_error} with randomly drawn input parameters assuming that all input parameters are normally distributed and uncorrelated, except for the theoretical error model parameters $w_{LCOPE}, w_{\alpha_s}^{(p)}, w_{n\text{-part}}^{(p)}$ which are uniformly distributed.
\begin{table}[H]    
    \centering
    \begin{tabular}{|c|c|c|}
    \hline
         Parameters & Values  & Ref. \\
         \hline \hline
          \multirow{7}{*}{Decay constants} & $f_B = 190.0 (1.3)$ MeV & \cite{FlavourLatticeAveragingGroupFLAG:2021npn} \\
          & $  f_\pi = 130.2 (8)$ MeV & \cite{FlavourLatticeAveragingGroupFLAG:2021npn}  \\
          & $f_K = 155.7 (7)$ MeV& \cite{FlavourLatticeAveragingGroupFLAG:2021npn} \\
          & $f_D = 212.0 (7) $  MeV& \cite{FlavourLatticeAveragingGroupFLAG:2021npn}\\
          & $f_\rho = 213 (5)$  MeV& \cite{Bharucha:2015bzk}\\
          & $f_{K*} = 204 (7)$ MeV& \cite{Bharucha:2015bzk} \\
          & $ f_{D*} =242 (20)$ MeV & \cite{Gelhausen:2013wia}\\
        \hline
        QCD coupling &  $\alpha_s(\mu = m_Z) = 0.1180 $ & \cite{Workman:2022ynf} \\
        \hline
          \multirow{3}{*}{Quark masses}  & $m_c = \overline{m_c}(\mu = 2 \, \textrm{GeV}) = 1.10 \, \textrm{GeV}$ & \cite{Aebischer:2018bkb} \\ 
    & $m_s = \overline{m_s}(\mu = 1 \, \textrm{GeV}) = 0.121 \, \textrm{GeV}$ & \cite{Aebischer:2018bkb} \\
    & $m_u = m_d = 0 \, \textrm{GeV}$ & \\ \hline
    \multirow{3}{*}{LCDA parameters} & $\lambda_E^2=0.03 (2)$ GeV$^2$ & \cite{Nishikawa:2011qk}\\
    & $\lambda_H^2 = 0.06 (3)$  GeV$^2$ & \cite{Nishikawa:2011qk}  \\
    & $\lambda_B^{-1} = 2.72 \pm 0.66 \, \textrm{GeV}^{-1} $& \cite{Khodjamirian:2020hob}$^*$ \\ \hline
    \end{tabular}
    \caption{Values of input parameters. $^*$Our combination (see Section \ref{sec:limitations}). }
    \label{tab:input_param}
\end{table}

\subsection{Semiglobal QHD vs our approach} \label{sec:QHDvsUS}
We begin the presentation of our numerical results by comparing the value of $\widetilde{\Pi}_{F}^{(p)}(q^2 = 0)$ and the prediction of $F(q^2 = 0)$ obtained by applying the semiglobal QHD. The purpose of this comparison is to verify that $\widetilde{\Pi}_F^{(p)}$ and $R_F$ behave as expected in Section \ref{sec:method}, and to estimate numerically the value of $-k^2/p$ for which the convergence regime starts.

For this check, we choose to study $F = V^{B \to K^*}$, for which we established that the DSR in Eq.~\eqref{eq:DSR} yields a reasonable effective threshold. We set $k^2=-20$ GeV$^2$ and vary $p$ such that $-k^2/p$ spans the range $\left[ 0.57, 1.53\right]$ GeV$^2$. Using the statistical routine described in Section \ref{sec:input} we obtain the central value and $68 \%$ C.L. interval for $\widetilde{\Pi}_{V^{B \to K^*}}^{(p)}$ as well as $R_F$ and $V^{B \to K^*}$ estimated using semiglobal QHD. The results are presented in Table \ref{tab:QHD}. 
\begin{table}[H]
   \centering
    \resizebox{\columnwidth}{!}{
       \begin{tabular}{|c||c|c|c|c|c|c|c|}
       \hline
           $-k^2/p$ ($\textrm{GeV}^2$) & $20/13 $ & $20/16$ & $20/20 $ & $20/25$ & $20/30$ & $20/35$ \\ \hline
            $R_{V^{B \to K^*}}$ & $0.50_{-0.13}^{+0.15}$ & $0.37_{-0.11}^{+0.11}$ & $0.25_{-0.08}^{+0.08}$ & $0.15_{-0.04}^{+0.05}$& $0.08_{-0.02}^{+0.03}$& $0.03_{-0.01}^{+0.01}$\\ \hline
            $V^{B \to K^*}$ & $0.43_{-0.19}^{+0.24}$ & $0.45_{-0.20}^{+0.25}$& $0.47_{-0.22}^{+0.28}$ & $0.49_{-0.21}^{+0.30}$ & $0.52_{-0.23}^{+0.31}$& $0.53_{-0.24}^{+0.35}$\\ \hline
            $\widetilde{\Pi}_{V^{B \to K^*}}^{(p)}$ & $0.94_{-0.33}^{+0.37}$ & $0.84_{-0.31}^{+0.36}$ & $0.72_{-0.30}^{+0.35}$& $0.64_{-0.26}^{+0.34}$ & $0.60_{-0.25}^{+0.32}$ & $0.55_{-0.25}^{+0.35}$  \\ \hline
            $s_0$(GeV$^2$) & $1.43_{-0.012}^{+0.026}$ & $1.48_{-0.01}^{+0.03}$ & $1.56_{-0.02}^{+0.04}$& $1.68_{-0.03}^{+0.06}$ & $1.87_{-0.05}^{+0.1}$ & $2.21_{-0.10}^{+0.25}$ \\ \hline
       \end{tabular}
       }
        \caption{Central values and $68 \%$ C.L. intervals of $\widetilde{\Pi}_{V^{B \to K^*}}^{(p)}$ ($q^2=0$). The corresponding $R_{V^{B \to K^*}}$, $V^{B \to K^*}$ and $s_0$  are estimated using DSR and semiglobal QHD.}
       \label{tab:QHD}
   \end{table} 
Focusing on the central values of our predictions, we first note that --as expected-- ${R_{V^{B \to K^*}} \to 0}$ and $\widetilde{\Pi}^{(p)}_{V^{B \to K^*}} \to V^{B \to K^*}$ when $-k^2/p$ is small enough, typically around $-k^2/p \sim 0.5$ GeV$^2$. Interestingly, this corresponds to the lower end of the Borel window used in the literature, indicating that our current knowledge of the LCOPE for $B$-meson LCSR may allow us to circumvent the semiglobal QHD. 
When approaching the convergence, the total uncertainty grows when $-k^2/p$ decreases which is also expected given our model for the theoretical error devised in Section \ref{sec:estimation}. 

Before pursuing the numerical analysis of our results, we comment on the size of $R_F$ estimated with semiglobal QHD at high $-k^2/p$. It is known that in this regime the spectral density integral $R_F$ is poorly suppressed, and in the literature the upper end of the Borel window is chosen to limit the impact of errors coming from the estimate of $R_F$. We find that for $M^2 \sim -k^2/p = 1.53$ GeV$^2$, $R_F/\widetilde{\Pi}^{(p)}_F = 53 \%$, which we argue is too large given our poor knowledge of the $\textrm{Im}\Pi_{F,LCOPE}(s)$ at large $s$. We stress that this result is not due to the absence of Borel transformation (which could explain a poor suppression of the spectral density integral), since the Borel transformation converges for $-k^2>5$GeV$^2$ at $-k^2/p \sim 1$. 

When using the semiglobal QHD, to fulfill the criteria $R_F / \widetilde{\Pi}^{(p)}_{F} < 30 \%$ we find that one should take $-k^2/p \sim M^2 \lesssim 1$ GeV$^2$. This restriction of the Borel window could have serious phenomenological consequences since, as can be seen in Table \ref{tab:QHD}, the variation of the form factor in the Borel window between $-k^2/p = 1.5$ GeV$^2$ and $-k^2/p = 1$ GeV$^2$ is of the order of $10 \%$. This observation is particularly relevant in the context of $B$ anomalies since a $10\%$ decrease of $f_+^{B \to K}(q^2=0)$ would bring the SM prediction of e.g. BR$(B^+ \to K^+ \mu \mu)$ closer to its experimental determination. 

\subsection{Prediction and upper limits on local form factors}

\begin{figure}
    \centering
    \includegraphics{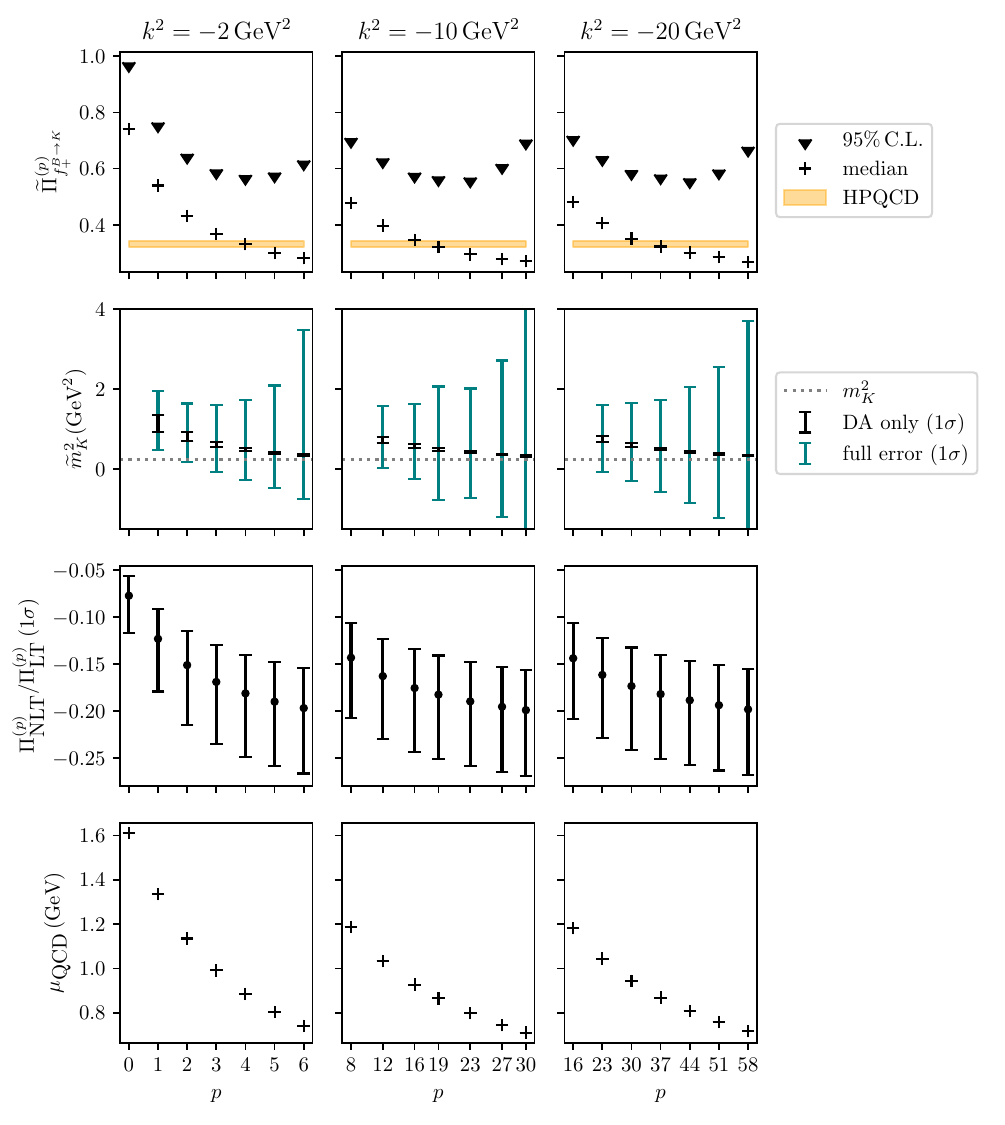}
    \caption{Numerical results for $k^2 = -2,-10,-20 \, \textrm{GeV}^2$ around the optimal value of $p$. \vspace{0.1cm} \\ 
        - $ 1^{\textrm{st}}$ row: $95^{\textrm{th}}$ percentile and median value of $\widetilde{\Pi}_{f_+}^{B \to K}$, $N=6000$. The orange bands represent the $1\sigma$ interval predicted by the HPQCD collaboration \cite{Parrott:2022rgu}.  \\
        - $2^{\textrm{nd}}$ row: blue (black) error bars are $1 \sigma$ intervals including all errors (parametric error only). Dotted line is experimental value of $m_K^2$. \\
        - $3^{\textrm{rd}}$ row: $1\sigma$ intervals of the ratio of next-to-leading twist over leading twist contributions. \\
        - $4^{\textrm{th}}$ row: energy scale used for the estimation of the error from missing radiative corrections.}
    \label{fig:fplusBtoK}
\end{figure}

We present results for $-k^2 = 2, 10, 20$ GeV$^2$ which we find are representative values. We choose $-k^2 = 2$ GeV$^2$ as the minimal value which fulfills the QCD perturbativity condition \eqref{eq:perturbativity1}. For each $-k^2$ we increase $p$ until the error in $\widetilde{\Pi}_{F,\, LCOPE}^{(p)}$ starts diverging. We then go back and sample a large number of points for a few values of $p$ before the divergence. We illustrate this method in Fig. \ref{fig:fplusBtoK} for the $f_+^{B \to K}$ form factor, and also include the predictions from the mass sum rule and other relevant quantities. Similar figures are given in Appendix \ref{appendix:extrafigures} for other form factors (see Figs. 4-6).

At $q^2=0$, we restrict ourselves to a minimal basis for the form factors: $f_+^{B \to P}$, $f_T^{B \to P}$ for pseudoscalar final mesons, and $V^{B \to V}$, $A_1^{B \to V}$, $A_2^{B \to V}$, $T_1^{B \to V}$ and $T_{23}^{B \to V}$ for vector final mesons (see Appendix \ref{appendix:FFdef} for their definitions).  We present our results for upper bounds and predictions in Tables \ref{tab:resultsBtoP} and \ref{tab:resultsBtoV} for $B \to \pi, \rho, K^{(*)}$. The optimal values for the pair $(k^2, p)$ are chosen to obtain the most stringent upper limit at the $95 \%$ confidence level for the considered form factor. We also present the median value and $1\sigma$ interval of the sampled $\widetilde{\Pi}_{F,\, LCOPE}^{(p)}$ at the said optimal pair. For all three values of $k^2$, the optimal $p$ is sought in the interval where the scale $\mu_{\text{QCD}} \geq 0.75$ GeV and the twist ratio is below $20 \%$. We also include in the tables an estimation of $R_F$ using semiglobal QHD to test its sign and magnitude. The result one would obtain from the usual LCSR method is simply $\widetilde{\Pi}_F^{(p)}$ - $R_F$. For $B \to \rho, K^*$, we solve the daughter sum rules Eq.~\eqref{eq:DSR} to determine the effective thresholds, while for $B \to \pi, K$ we use the same effective thresholds as in Ref. \cite{Gubernari:2018wyi}. 

For $B\to D^{(*)}$, we do not obtain viable results for two reasons. First, the NLT contribution is much larger, due to the large mass of the charm quark, which makes the error large before reaching the convergence. Secondly, the daughter sum rules do not work, preventing us from evaluating $R_F$ and checking its positivity. 

The central values of $\widetilde{\Pi}_F^{(p)}$ can be used as indications for the predictions of the form factors assuming the convergence of the sum rule. We discuss convergence further in Section \ref{sec:convergence}. We find that all predictions are compatible with the literature albeit the uncertainties are rather large. The large error in our predictions and the elevated values of the upper limits are mainly due to the large uncertainties coming from the estimation of the error from the truncation of perturbative QCD corrections and to a lesser extent to the truncation of the LCOPE. This demonstrates the necessity of a more accurate assessment of the radiative corrections and the higher twist contributions, which would allow us to go at a larger number of derivation $p$ and get closer to the convergence and/or reduce the theoretical error. 

From the three models introduced in Appendix \ref{appendix:B_LCDAs}, the results in Tables \ref{tab:resultsBtoP} and \ref{tab:resultsBtoV} were obtained with the exponential model. It yields smaller yet compatible values compared to the local duality models A and B, with relative differences ranging from $0$ to $10 \%$. These models are asymptotically identical at low momentum, hence in the limit $p\to \infty$ there is no difference between them. 

\begin{table}[h]
    \centering
    \begin{tabular}{|c||c|c||c|c|c|c|}
    \hhline{-||--||----}
         Form factor& $-k^2/p $ & $R_F(p,k^2)$ & \makecell{Upper limit \\ @ $95\%$ C.L.} & $\widetilde{\Pi}_F^{(p)}$ $(1\sigma)$ & Literature & Ref.\\ \hhline{=::==::====}
          $f_+^{B\to \pi}$ & $2/6$ &  $0.02_{-0.01}^{+0.01}$ &  $0.38$ & $0.17_{-0.10}^{+0.13}$ & \makecell{$0.21(7)$ \\ $0.191(73)$  \\ $0.301(23)$ \\ $0.297(30)$ } & \makecell{\cite{Gubernari:2018wyi}$^\dagger$ \\ \cite{Cui:2022zwm} \\ \cite{Khodjamirian:2017fxg} \\ \cite{Leljak:2021vte} } \\ \hhline{-||--||----}
          $f_T^{B\to \pi}$ & $2/5$ & $0.016_{-0.006}^{+0.007}$ &  $0.32$ & $0.17_{-0.08}^{+0.09}$ & \makecell{$0.19(7)$ \\ $0.222(78)$ \\ $0.273(21)$ \\ $0.293(28)$ }& \makecell{\cite{Gubernari:2018wyi}$^\dagger$ \\ \cite{Cui:2022zwm}  \\ \cite{Khodjamirian:2017fxg} \\ \cite{Leljak:2021vte} } \\ \hhline{=::==::====}
          $f_+^{B\to K}$ & $10/19$ &  $0.03_{-0.01}^{+0.01}$  &$0.57$ & $0.32_{-0.12}^{+0.15}$ &  \makecell{$0.332(12)$ \\ $0.27(8)$ \\ $0.325(85)$ \\ $0.395(33)$ } & \makecell{\cite{Parrott:2022rgu} \\ \cite{Gubernari:2018wyi}$^\dagger$ \\ \cite{Cui:2022zwm} \\ \cite{Khodjamirian:2017fxg}  }\\ \hhline{-||--||----}
          $f_T^{B\to K}$ & $10/8$ & $0.04_{-0.06}^{+0.02}$ &  $0.46$ & $0.34_{-0.07}^{+0.08}$ & \makecell{$0.332(21)$ \\ $0.25(7)$ \\ $0.381(27)$ \\ $0.381(97)$} & \makecell{\cite{Parrott:2022rgu} \\ \cite{Gubernari:2018wyi}$^\dagger$ \\ \cite{Khodjamirian:2017fxg} \\ \cite{Cui:2022zwm}} \\ \hhline{-||--||----}
        \end{tabular}
    \caption{Upper limits at the $95 \%$ confidence level and central value of $\widetilde{\Pi}_F^{(p)}$ for $B \to \pi, K$. We include the corresponding values of $-k^2$ (in GeV$^2$) and $p$ as well as an estimate of $R_F(p,k^2)$ using semiglobal quark-hadron duality with the effective thresholds $s_0$ of \cite{Gubernari:2018wyi}. $^\dagger$The authors of \cite{Gubernari:2018wyi} have recommended not to use their results, yet they are of interest since we follow a similar calculation (see text for more details).}
    \label{tab:resultsBtoP}
\end{table}

\begin{table}[h]
    \centering
    \begin{tabular}{|c||c|c|c||c|c|c|c|c|}
    \hhline{-||---||----}
        Form factor & $-k^2/p $ & $R_F(p)$ & $s_0$ (GeV$^2$) & \makecell{Upper limit \\ @ $95\%$ C.L.} & $\widetilde{\Pi}_F^{(p)}$ $(1\sigma)$ & Literature & Ref. \\ \hhline{=::===::====}
          $V^{B\to \rho}$ & $ 20/44 $ & $0.05_{-0.02}^{+0.03}$ & $1.35_{-0.06}^{+0.06}$ & $0.82$ & $0.34_{-0.18}^{+0.28}$ &\makecell{$0.27(14)$ \\ $0.327^{+0.204}_{-0.135}$ \\ $0.327(31)$} & \makecell{ \cite{Gubernari:2018wyi} \\ \cite{Gao:2019lta} \\ \cite{Bharucha:2015bzk}} \\ \hhline{-||---||----}
          $A_1^{B\to \rho}$ & $20/44$ & $0.04_{-0.02}^{+0.02}$ & $1.36_{-0.05}^{+0.07}$ & $0.63$ & $0.26_{-0.13}^{+0.21}$& \makecell{$0.22(10)$ \\ $0.249^{+0.155}_{-0.103}$ \\ $0.262(26)$} & \makecell{ \cite{Gubernari:2018wyi} \\ \cite{Gao:2019lta} \\ \cite{Bharucha:2015bzk}} \\ \hhline{-||---||----}
          $A_2^{B\to \rho}$ & $20/37$ & $0.07_{-0.03}^{+0.05}$ & $1.22_{-0.03}^{+0.04}$ & $0.70$ & $0.26_{-0.14}^{+0.25}$ & \makecell{$0.19(11)$} & \makecell{\cite{Gubernari:2018wyi}} \\ \hline
          $T_1^{B\to \rho}$ & $20/37$ & $0.08_{-0.03}^{+0.04}$ & $1.23_{-0.04}^{+0.04}$ & $0.72$ & $0.33_{-0.16}^{+0.22}$ & \makecell{$0.24(12)$ \\ $0.272(26)$} & \makecell{\cite{Gubernari:2018wyi} \\ \cite{Bharucha:2015bzk}} \\ \hhline{-||---||----}
          $T_{23}^{B\to \rho}$ &  $2/3^{**}$ & - & - & $0.93$ & $0.68_{-0.12}^{+0.14}$ & \makecell{$0.56(15)$ \\ $0.747(76)$} & \makecell{\cite{Gubernari:2018wyi} \\ \cite{Bharucha:2015bzk}} \\ \hhline{=::===::====}
          $V^{B\to K^*}$ & $20/30$ & $0.08_{-0.02}^{+0.03}$ & $1.87_{-0.04}^{+0.09}$ & $1.1$ & $0.58_{-0.25}^{+0.34}$& \makecell{$0.33(11)$ \\ $0.419^{+0.245}_{-0.157}$ \\ $0.341(36)$} & \makecell{\cite{Gubernari:2018wyi} \\ \cite{Gao:2019lta} \\ \cite{Bharucha:2015bzk}}\\ \hhline{-||---||----}
          $A_1^{B\to K^*}$ & $10/16$ & $0.04_{-0.01}^{+0.02}$ & $2.05_{-0.08}^{+0.14}$ & $0.88$ & $0.45_{-0.19}^{+0.25}$& \makecell{$0.26(8)$ \\ $0.306^{+0.180}_{-0.115}$ \\ $0.269(29)$} & \makecell{\cite{Gubernari:2018wyi} \\ \cite{Gao:2019lta} \\ \cite{Bharucha:2015bzk}}\\ \hhline{-||---||----}
          $A_2^{B\to K^*}$ & $20/31$ & $0.04_{-0.02}^{+0.02}$ & $2.03_{-0.03}^{+0.11}$ & $0.96$ & $0.42_{-0.21}^{+0.30}$& \makecell{$0.24(9)$} & \makecell{\cite{Gubernari:2018wyi}} \\ \hhline{-||---||----}
          $T_1^{B\to K^*}$ & $10/16$ & $0.04_{-0.01}^{+0.01}$& $2.05_{-0.07}^{+0.16}$ & $1.0$ & $0.50_{-0.22}^{+0.28}$& \makecell{$0.29(10)$ \\ $0.361^{+0.211}_{-0.135}$ \\ $0.282(31)$} & \makecell{\cite{Gubernari:2018wyi} \\ \cite{Gao:2019lta} \\ \cite{Bharucha:2015bzk}} \\ \hhline{-||---||----}
          $T_{23}^{B\to K^*}$ & $20/26^{**}$ & - & - & $1.2$& $0.87_{-0.20}^{+0.22}$ & \makecell{$0.81(11)$ \\ $0.793^{+0.402}_{-0.258}$ \\ $0.668(83)$} & \makecell{\cite{Gubernari:2018wyi} \\ \cite{Gao:2019lta} \\ \cite{Bharucha:2015bzk}} \\ \hhline{-||---||----}
    \end{tabular}
    \caption{Upper limits at the $95 \%$ confidence level and central value of $\widetilde{\Pi}_F^{(p)}$ for $B \to \rho,K^*$. We include the corresponding values of $-k^2$ (in GeV$^2$) and $p$ as well as an estimate of $R_F(p,k^2)$ using semiglobal quark-hadron duality. The effective thresholds $s_0$ are determined by solving the daughter sum rule \eqref{eq:DSR}. $^{**}$ $-k^2/p$ is taken from the sum rule for $T_{23}^B$.}
    \label{tab:resultsBtoV}
\end{table}

\subsection{Convergence of the sum rule} \label{sec:convergence}

The first criterion of convergence discussed in Section \ref{sec:method} is the independence of the prediction with respect to $k^2$ and $p$. At this stage, we find that it is the case within uncertainties, but this is rather meaningless given how large the latter are. The second criterion of convergence we examined is the convergence of $\widetilde{m}_M^2 \to m_M^2$. Similarly to the convergence of the form factor, taken at face value, the total uncertainty on mass predictions is very large, and thus, mass predictions are not very useful at this stage to characterise convergence. They are dominated by our QCD error estimation, and computing the radiative corrections should reduce this uncertainty by a fair amount and allow us to check the convergence of the sum rule.

So far, the behaviour we have described was expected. However, we make an interesting observation in the mass sum rule. As shown in Fig.\ref{fig:fplusBtoK}, the value of $\widetilde{m}^2_K$ including parametric errors only becomes remarkably close to the squared meson mass $m_K^2$ as $p$ increases, with very small parametric uncertainties. This is also true for $f_T^{B \to K}$ and for all the form factors for $B \to \rho$ and $B \to K^*$ transitions as shown in Appendix \ref{appendix:extrafigures}. 
We checked numerically that there is no correlation between the input parameter $m_M$ and the calculated $\widetilde{m}_M$. It is a truly surprising finding given how large the estimated error is. At this stage, this appears to be a numerical coincidence which has two possible explanations. Either the actual radiative corrections and higher twists contributions are small or, more probably, they cancel out in the ratio \eqref{eq:lim_m2}. In any case, such a quick and accurate convergence seems to be a sign that $R_F \ll F(q^2)$ for a relatively large $-k^2/p$ where the radiative corrections (although potentially large) are calculable perturbatively for $B \to K^{(*)}, \rho$. 

The case of $B \to \pi$ is different. In $B$-meson LCSRs, the $SU(3)_f$ breaking effects are very small at leading order in QCD, hence the mass sum rule yields $\widetilde{m}_K^2 \approx \widetilde{m}_\pi^2$. We expect that the radiative corrections break strongly $SU(3)_f$ to adjust the predicted mass of the pion. Notably, the corrections should be larger for $B \to \pi$ since the first resonance of the pion spectral density is lighter than the first resonance of the kaon, $s_{cont}^\pi < s_{cont}^K$, and thus, $R_{F^{B \to \pi}}$ converges more slowly than $R_{F^{B \to K}}$.

As a check, the estimated values of $R_F$ including QCD and LCOPE truncation error are also reported for each form factor. Assuming semiglobal QHD, it allows us to infer whether we can impose an upper limit by checking the positivity of $R_F$. It also gives an indication of how far we are from the convergence of the sum rule since we expect $R_F$ to go to zero in this limit. We find that for every prediction $R_F$ is much smaller than the central value of the form factors, suggesting that we are close to convergence.

From the Tables \ref{tab:resultsBtoP} and \ref{tab:resultsBtoV}, one can see that the ratios $-k^2/p$ taken to establish the upper bounds are around or below $0.5$ GeV$^2$, the canonical lower limit for the Borel window. There are two reasons to explain this. First, in our approach we allow larger (yet conservatively estimated) truncation errors in order to better suppress the spectral density integral $R_F$. Then, in our expression of the average virtuality in the correlator \eqref{eq:avgs}, there is a disconnect between $-k^2/p$ and $\langle s \rangle$ for small Borel parameters, e.g. in Fig. \ref{fig:fplusBtoK}, the point $k^2 = -2$ GeV$^2$ and $p=6$ corresponds to $\langle s \rangle \approx 0.56$ GeV$^2$, larger than the expected $-k^2/p = 0.33$ GeV$^2$. This reduces the estimated size of radiative correction for small Borel parameters.

\subsection{Correlation between form factors}

\begin{figure}[h]
  \centering
  \subfloat{\includegraphics[width=0.5\textwidth]{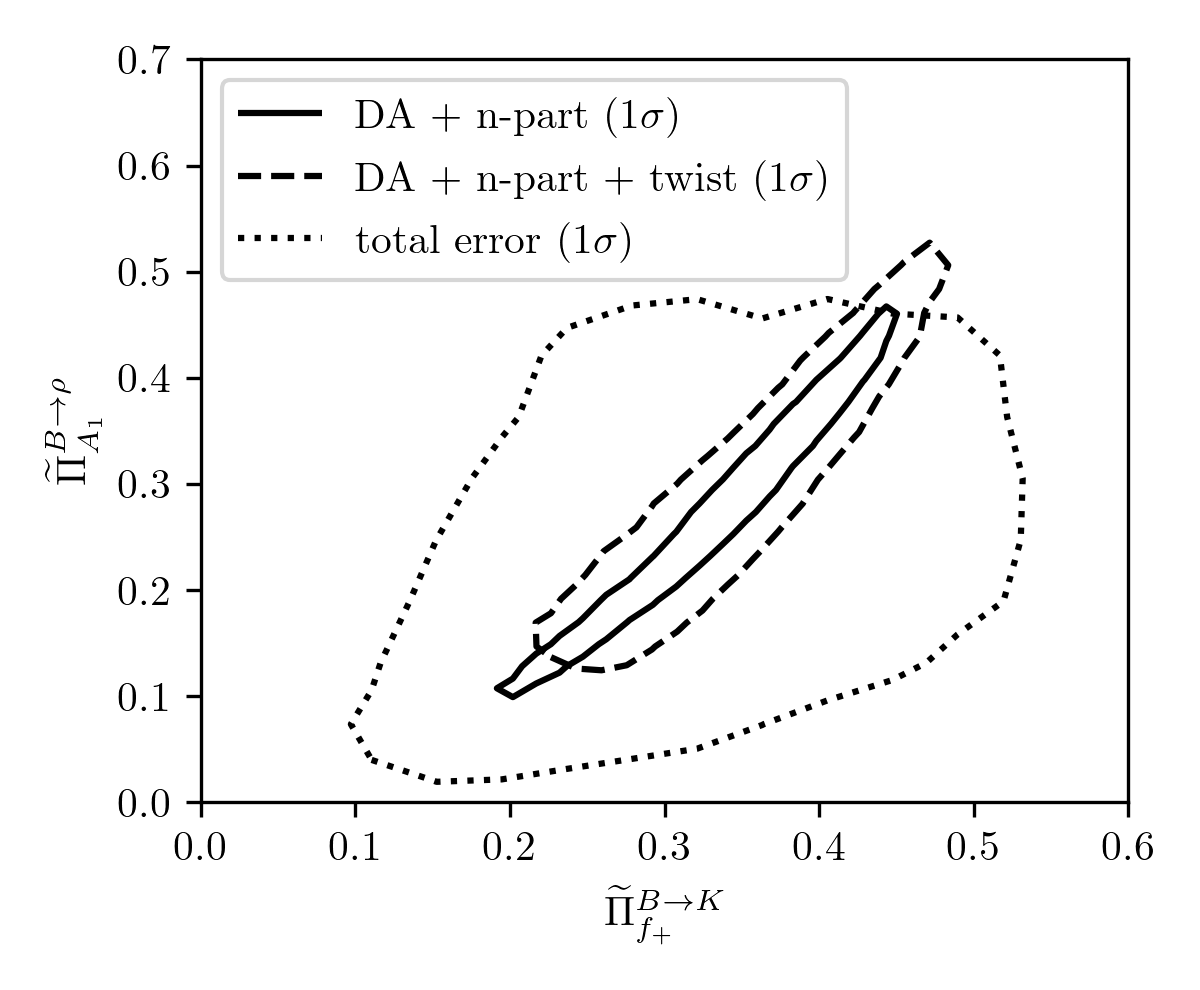}\label{fig:f1}}
  \hfill
  \subfloat{\includegraphics[width=0.5\textwidth]{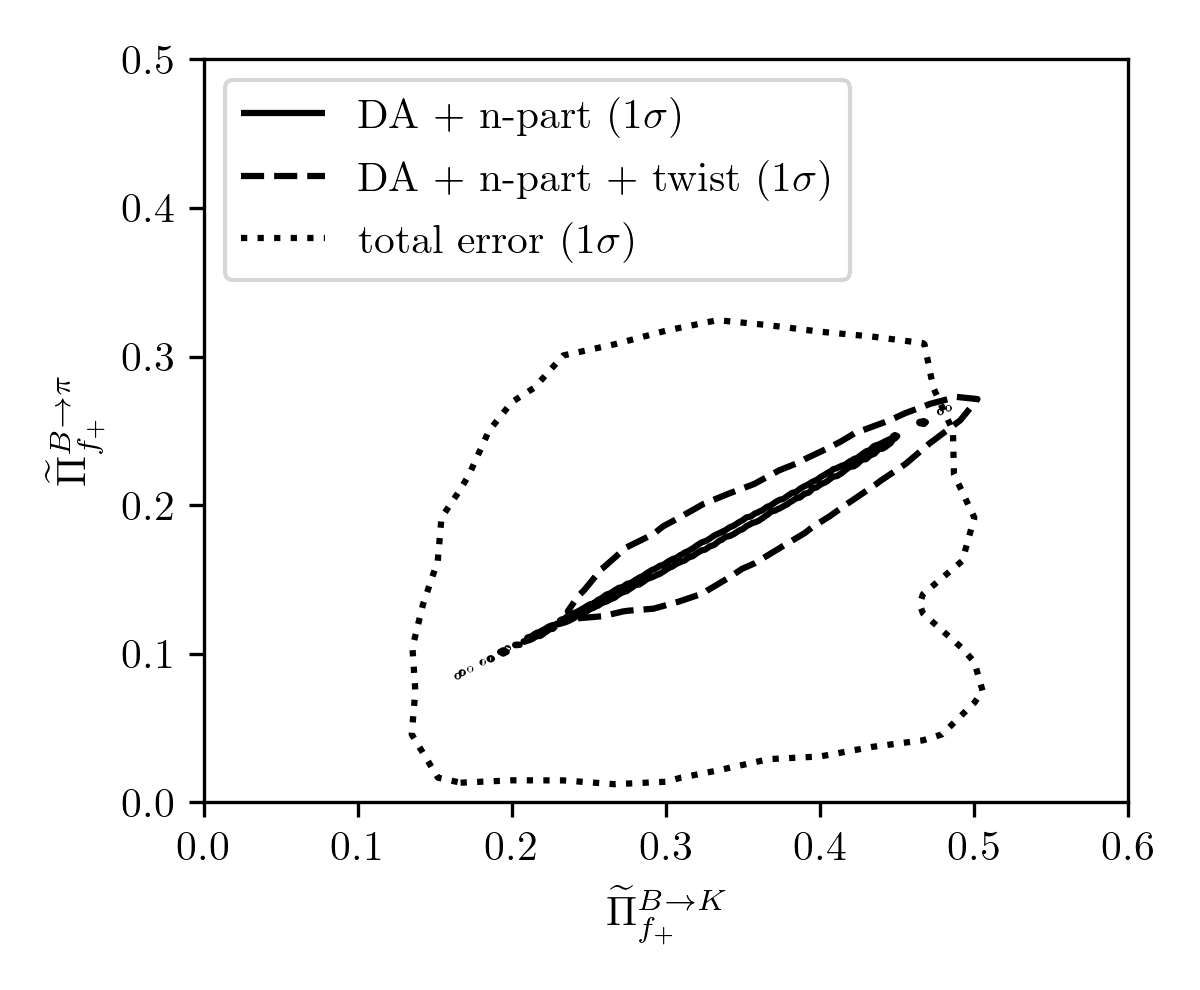}\label{fig:f2}}
  \caption{Correlated predictions of $\widetilde{\Pi}_{f_+}^{B\to K}$-$\widetilde{\Pi}_{A_1}^{B\to \rho}$ and $\widetilde{\Pi}_{f_+}^{B\to K}$- $\widetilde{\Pi}_{f_+}^{B\to \pi}$ at the $1 \sigma$ confidence level. See Tables \ref{tab:resultsBtoP} and \ref{tab:resultsBtoV} for the values of $-k^2/p$. The solid contour only includes parametric errors (dominated by the LCDA) and the error from truncating the $n$-particle expansion of the correlator. The dashed contour includes the aforementioned errors as well as the error from the truncation of the twist expansion of the correlator. The dotted contour includes all previous errors as well as the error coming from the unaccounted radiative corrections from QCD.}
  \label{fig:FFs_correlation}
\end{figure}

Since the core result of $B$-meson LCSRs relies on the same shared LCDA parameters for all form factors, we expect this calculation to yield strongly correlated predictions. Moreover, for a given form factor, the correlation function is the same for different processes up to the interpolating quark mass. We show an example of this effect in Fig.\ref{fig:FFs_correlation}.
The correlation between form factors predicted using $B$-meson LCSRs is usually hampered by the determination of the Borel parameter and effective threshold which have a relatively large uncertainty and are \textit{a priori} uncorrelated between different decays. Since we avoid semiglobal QHD in our procedure, we can recover a strong correlation between all predicted form factors, as long as the uncertainty coming from $B$-meson LCDAs dominates. A large part of the uncertainty in our current calculation originates from perturbative QCD, which dilutes the correlations for now.

In Fig.~\ref{fig:FFs_correlation}, we show the $1\sigma$ confidence levels of our predictions in the planes $\widetilde{\Pi}_{f_+}^{B\to K} - \widetilde{\Pi}_{A_1}^{B\to \rho}$ and $\widetilde{\Pi}_{f_+}^{B\to K} - \widetilde{\Pi}_{f_+}^{B\to \pi}$ breaking down the different sources of error. When accounting for the total error, there is virtually no correlation between our predictions, which is expected because of the large and uncorrelated QCD error. Removing the latter yields a strong correlation between our predictions, even when accounting for LCOPE truncation errors. Keeping only parametric errors, we see even stronger correlations, reaching almost $100\%$ correlation between $f_+^{B \to K}$ and $f_+^{B \to \pi}$ which is expected at QCD LO for $B$-meson LCSR. Even without improving our current knowledge of $B$-meson distribution amplitudes, calculating the QCD radiative corrections to the correlator would significantly increase the statistical correlation between all predicted form factors. While the absolute size of the error would still be comparable to other calculations employing $B$-meson LCDAs, this correlation can have strong phenomenological implications. 

\section{Conclusion}

In this article, we propose a strategy to predict form factors using LCSRs without the semiglobal QHD approximation, and thus without relying on the determination of an effective threshold and a window for the Borel parameter. Our approach consists of evaluating conservatively the truncation error in the correlator, and taking $-k^2/p \approx M^2$ to be as small as possible in order to suppress the spectral density integral $R_F$. We introduce two tests of the suppression of $R_F$ which do not rely on semiglobal QHD, namely the $k^2$ and $p$ independences of the predicted form factors, and the convergence of the mass sum rule toward the physical value of the squared meson mass $\widetilde{m}_M^2 \to m_M^2$. 
Our approach reduces greatly the systematic uncertainty associated to the LCSR method. However, this improvement comes at the cost of a larger dependence on higher-order perturbative corrections and higher twists in the LCOPE. This trade-off is advantageous, since these corrections are calculable. The other notable advantage of this method is that in $B$-meson LCSRs it predicts strongly correlated form factors between different processes.

Using a conservative model of the error in $\widetilde{\Pi}_F^{(p)}$, we find that $-k^2/p$ can be taken slightly below the Borel window $\left[0.5, 1.5\right]$ GeV$^2$ used in the literature \cite{Khodjamirian:2006st,Gubernari:2018wyi}, thanks to the fact that a direct assessment of the average virtuality $\langle s \rangle$ yields higher values than the approximation $-k^2/p \approx \langle s \rangle$ and thus hints at smaller expected radiative corrections. At this stage, an assessment of the convergence of the sum rule in this regime is hampered by the relatively large truncation error which affects both $\widetilde{\Pi}_F^{(p)}$ and $\widetilde{m}_M^2$. However, we find that without the modeled truncation error, $\widetilde{m}_M^2$ converges remarkably fast and accurately to the physical values of $m_M^2$ for $M = K^{(*)},\rho$. Our understanding of this observation is that $R_F$ converges quickly to zero, and the error due to the truncated contributions in pQCD and LCOPE cancels out in the ratio in the mass sum rule provided in Eq.~\eqref{eq:lim_m2}. In this case, computing the NLO QCD correction could allow us to demonstrate the convergence in the region $-k^2/p \approx 0.5-0.3$ GeV$^2$. In addition, the knowledge of these radiative corrections would lead to highly correlated predictions of the form factors. 

Interestingly, the same method can be applied to LCSRs with light-meson LCDAs and should prove to perform better. Indeed, the light-meson LCDAs are generally better known, with smaller parametric uncertainties \cite{Duplancic:2008ix}. They do not exhibit UV divergences similar to the one mentioned in Section \ref{sec:pole}. Furthermore, for the $B \to \pi$ and $B \to K$ transitions, the contributions from the highest known twists have been shown to be negligible \cite{Rusov:2017chr, Khodjamirian:2017fxg}, hence indicating that the LCOPE should be under control at small $-k^2/p$ in our approach. 
For each form factor considered in this work, the NLO QCD corrections have already been calculated in LCSRs with light-meson LCDAs
\cite{Ball:2004ye, Duplancic:2008ix, Bharucha:2015bzk, Khodjamirian:2023wol}, at least for the leading twist contributions. It would be interesting to compare the results of this method applied with light-meson LCDAs to the ones already obtained with the usual procedure. Light-meson LCDAs also present the advantage of enabling exploration of higher values of $q^2$, surpassing the restriction $q^2 < m_1^2$.

\paragraph{Acknowledgement:} 
The authors are grateful to A. Khodjamirian, T. Hurth, D. van Dyk and M. Reboud for useful comments.
This work was supported in part by the National Research Agency (ANR) under project ANR-21-CE31-0002-01, and the work of A.C. was supported in part by the Italian Ministry of University and Research under grant no. 2022N4W8WR.

\newpage
\appendix
\section{Definition of local hadronic form factors} \label{appendix:FFdef}

The relevant form factors for the $B \to P$ transitions, with $P = \pi, K , \bar{D}$ are $f^{B \to P}_0$, $f^{B \to P}_+$ and $f^{B \to P}_T$. We define:
\begin{equation}
\begin{aligned}
\left\langle P(k)\left|\bar{q}_1 \gamma^\mu b\right| B(p_B)\right\rangle & =\left[(p_B+k)^\mu-\frac{m_B^2-m_P^2}{q^2} q^\mu\right] f_{+}^{B \rightarrow P}+\frac{m_B^2-m_P^2}{q^2} q^\mu f_0^{B \rightarrow P}\,, \\
\left\langle P(k)\left|\bar{q}_1 \sigma^{\mu \nu} q_\nu b\right| B(p_B)\right\rangle & =\frac{i f_T^{B \rightarrow P}}{m_B+m_P}\left[q^2(p_B+k)^\mu-\left(m_B^2-m_P^2\right) q^\mu\right] \,.
\end{aligned}
\end{equation}
For $B \to V$ transitions, with $V = \rho,  K^*, \bar{D}^*$, we consider $V^{B \rightarrow V}$, $A_0^{B \rightarrow V}$, $A_1^{B \rightarrow V}$, $A_2^{B \rightarrow V}$, $T_1^{B \rightarrow V}$, $T_2^{B \rightarrow V}$ and $T_3^{B \rightarrow V}$, which can be defined with: 
\begin{equation}
\begin{aligned}
\left\langle V(k, \eta)\left|\bar{q}_1 \gamma^\mu b\right| B(p_B)\right\rangle = & \epsilon^{\mu \nu \rho \sigma} \eta_\nu^* p_{B\rho} k_\sigma \frac{2 V^{B \rightarrow V}}{m_B+m_V}, \\
\left\langle V(k, \eta)\left|\bar{q}_1 \gamma^\mu \gamma_5 b\right| B(p_B)\right\rangle = & i \eta_\nu^*\left[q^\mu q^\nu \frac{2 m_V}{q^2}A_0^{B \rightarrow V} + \left(g^{\mu \nu} - \frac{q^\mu q^\nu}{q^2}\right)\left(m_B+m_V\right) A_1^{B \rightarrow V} \right.\\
& -\left(\frac{(p_B+k)^\mu q^\nu}{m_B+m_V} - \frac{q^\mu q^\nu}{q^2}(m_B - m_V)\right)A_2^{B \rightarrow V}  \biggr] , \\
\left\langle V(k, \eta)\left|\bar{q}_1 i \sigma^{\mu \nu} q_\nu b\right| B(p_B)\right\rangle 
 = & - \epsilon^{\mu \nu \rho \sigma} \eta_\nu^* p_{B\rho} k_\sigma 2 T_1^{B \rightarrow V}, \\
\left\langle V(k, \eta)\left|\bar{q}_1 i \sigma^{\mu \nu} q_\nu \gamma_5 b\right| B(p_B)\right\rangle = & i \eta_\nu^*\left[\left(g^{\mu \nu}\left(m_B^2-m_V^2\right)-(p_B+k)^\mu q^\nu\right) T_2^{B \rightarrow V}\right. \\
 & + q^\nu\left(q^\mu-\frac{q^2}{m_B^2-m_V^2}(p_B+k)^\mu\right) T_3^{B \rightarrow V}\bigr]\,,
\end{aligned}
\end{equation}
where $p_B$ denotes the momentum of the $B$ meson, $k$ that of the light meson and $q$ the momentum transfer, while $\eta$ stands for the polarisation of the vector meson. We work in the $\epsilon_{0123} = +1$ convention. We also introduce the form factor combination $T_{23B}^{B \to V}$ as in \cite{Gubernari:2018wyi}, motivated by its easy extraction directly from a sum rule. It is defined as: 
\begin{equation}
    \begin{aligned}
        T_{23B}^{B \to V} = \frac{T_2^{B \to V}}{2} + \frac{1}{2} \bigg(\frac{q^2}{m_B^2 - m_V^2} - 1 \bigg) T_3^{B \to V}\,. \\
    \end{aligned}
\end{equation}

We work at $q^2 = 0$, at which there are extra relations between the different form factors:
\begin{equation}
\begin{aligned}
    &f_+^{B \to P} (q^2 = 0) = f_0^{B \to P} (q^2 = 0)\,,\\
    &T_1^{B \to V} (q^2 = 0) = T_2^{B \to V} (q^2 = 0)\,, \\
    &A_0^{B \to V} (q^2 = 0) = \frac{m_B + m_V}{2 m_V} A_1^{B \to V} (q^2 = 0) - \frac{m_B - m_V}{2  m_V}  A_2^{B \to V} (q^2 = 0)\,.
\end{aligned}
\end{equation}

We also define $T_{23}^{B \to V}$ as: 
\begin{equation}
    T_{23}^{B \to V} \equiv \frac{(m_B^2-m_V^2)(m_B^2 + 3 m_V^2 - q^2)T_2^{B \to V} - \lambda(q^2)T_ 3^{B \to V}}{8m_Bm_V^2(m_B - m_V)} \,,
\end{equation}
where $\lambda \equiv [(m_B + m_V)^2 - q^2][(m_B-m_V)^2-q^2]$ is the Källén function. At $q^2 = 0$ this reduces to:
\begin{equation}
     T_{23}^{B \to V}(q^2 = 0) = \frac{(m_B + m_V)}{4 m_B m_V^2}  \bigg(2 m_V^2 T_2^{B \to V}(q^2=0) + (m_B^2 - m_V^2) T_{23B}^{B \to V}(q^2=0) \bigg)\,.
\end{equation}
\section{$B$-meson distribution amplitudes} \label{appendix:B_LCDAs}
The non-local matrix elements in Eqs.~\eqref{2p} and \eqref{3p} are parameterised with $B$-meson LCDAs \cite{Grozin:1996pq,Braun:2017liq} as: 

\begin{equation}\label{2p-DAs}
    \begin{split}
    \bra{0}\bar{q_2}^\alpha(x)[x, 0] h_{v}^{\beta}(0) \ket{\bar{B}(v)}  & = \\ 
    -\frac{i f_B m_B}{4}\int_0^{+ \infty} d\omega e^{-i\omega v.x} & \Bigg\{(1 + \slashed{v})\Bigg(\left[ \phi_{+}(\omega) + x^2 g_{+}(\omega) \big) \right]  \\
    &- \frac{\slashed{x}}{2} \frac{1}{v x}\left[\big(\phi_+ - \phi_- \big)(\omega) + x^2  \big(g_+ - g_- \big)(\omega)\right]\Bigg) \gamma_5\Bigg\}^{\beta \alpha}\,, 
    \end{split}
\end{equation}

\begin{align}\label{3p-DAs}
    &\bra{0}\bar{q_2}^\alpha(x)[x, ux] G_{\lambda \rho}(ux)   [ux, 0] h_v^\beta(0) \ket{\bar{B}(v)}  = \nonumber \\
    & \frac{f_Bm_B}{4}\int_0^{+ \infty}d\omega_1 \int_0^{+ \infty}d\omega_2 e^{-i(\omega_1+u\omega_2)v.x} \Bigg\{(1 + \slashed{v})\Bigg[ (v_\lambda \gamma_\rho - v_\rho \gamma_\lambda)\big(\psi_A -\psi_V \big) - i \sigma_{\lambda \rho} \psi_V \nonumber \\
    & -\frac{1}{v.x}(x_\lambda v_\rho - x_\rho v_\lambda) X_A + \frac{1}{v.x}(x_\lambda \gamma_\rho - x_\rho \gamma_\lambda)\big(W + Y_A \big) -\frac{i}{v.x}\epsilon_{\lambda \rho \mu \nu} x^\mu v^\nu \gamma_5 \tilde{X}_A \nonumber \\
    & +\frac{i}{v.x}\epsilon_{\lambda \rho \mu \nu} x^\mu \gamma^\nu \gamma_5 \tilde{Y}_A -\frac{1}{(v.x)^2}(x_\lambda v_\rho - x_\rho v_\lambda) \slashed{x}W +\frac{1}{(v.x)^2}(x_\lambda \gamma_\rho - x_\rho \gamma_\lambda)  \slashed{x}Z\Bigg] \gamma_5\Bigg\}^{\beta \alpha}(\omega_1,\omega_2)\,.
\end{align}

The brackets $\left[ x, 0\right]$ and such denote Wilson lines that render the LCDAs gauge invariant. We work in the Fock-Schwinger gauge $x^\mu A^a\mu(x)\lambda^a/2=0$ where the Wilson lines are \textbf{1}, and adopt again the convention $\epsilon_{0123} = +1$.\\
While the 3-particle $B$-LCDAs basis in \eqref{3p-DAs} has the advantage of having simple Lorentz structures, for LCSRs it is more convenient to work in a basis of LCDAs with definite collinear twists: 
\begin{align}
    &\phi_3 = \psi_A - \psi_V \,, \nonumber\\
    &\phi_4 =  \psi_A + \psi_V  \,, \nonumber\\
    &\psi_4 =  \psi_A + X_A  \,, \nonumber\\
    &\tilde{\psi}_4 =  \psi_V - \tilde{X}_A \,,  \nonumber\\
    &\tilde{\phi}_5 =  \psi_A + \psi_V + 2 Y_A - 2 \tilde{Y}_A +2 W \,,\\
    &\psi_5 =  -\psi_A +X_A - 2 Y_A \,,\nonumber\\
    &\tilde{\psi}_5 =  -\psi_V - \Tilde{X}_A + 2 \tilde{Y}_A \,,\nonumber\\
    &\phi_6 =  \psi_A - \psi_V  + 2 Y_A + 2 \tilde{Y}_A + 2 W - 4 Z \,. \nonumber
\end{align}

For both 2-particle and 3-particle LCDAs we consider all three models given in \cite{Braun:2017liq}, namely the exponential and the two local-duality models: local duality A and local duality B. Only the Wandzura-Wilczek approximation was given for $g_-$ in \cite{Gubernari:2018wyi}, as the full LCDA depends on $\psi_5$, which also lacked a specified model. More recently models for the exponential model and the local-duality B one were given for $\psi_5$ in \cite{Gao:2019lta}. A general ansatz was later derived in \cite{Cui:2022zwm} that we use to obtain an expression for all three models. We check that we are in agreement with \cite{Gao:2019lta}. \\[1\baselineskip]
For the complete expression of $g_-(\omega)$, following \cite{Braun:2017liq} we start from the equation:
 \begin{equation}
    2 z^2 G_-(z) = - \Big[z \frac{d}{dz} - \frac{1}{2} + i z \Bar{\Lambda}  \Big] \Phi_-(z) - \frac{1}{2} \Phi_+(z) - z^2 \int_0 ^1 (1-u)du\Psi_5(z, uz)\,,
\end{equation}
where \begin{equation}
    G_-(z) = \int_0^{+\infty}d\omega e^{-iwz} g_-(\omega)\,,
\end{equation}
and \begin{equation}
    \Bar{\Lambda} = m_B - m_b\,.
\end{equation}
This yields: 
\begin{align}
    g_-(\omega) = &\hspace{0.2cm}\frac{1}{4} \int_0^{+\infty}d\rho (\Bar{\Lambda} - \rho) \text{Sign}(\omega - \rho) \phi_-(\rho) \nonumber \\
    & - \frac{1}{8} \int_0^{+ \infty} d\rho(\rho - \omega)\text{Sign}(\omega - \rho) \big[ \phi_+(\rho) - \phi_-(\rho) \big]\nonumber \\
    & -\frac{1}{2} \int_0^{\omega} d\omega_1 \int_0^1 du \frac{1-u}{u}\psi_5(\omega_1, \frac{\omega - \omega_1}{u})\,.
\end{align}
\\[1\baselineskip]

\textbf{Exponential Model:}\\[1\baselineskip]
\begin{align}
    &\psi_5(\omega_1, \omega_2) = -\frac{\lambda_E^2}{3 \omega_0^3} \omega_2 e^{-(\omega_1 + \omega_2) / \omega_0}\,,  \\
    & g_-(\omega) = \frac{1}{36 \omega_0^4} \Bigg\{e^{- \omega / \omega_0} \Big[27 \omega \omega_0^4 + (\lambda_E^2 - \lambda_H^2) \big(- \omega^3 + 3 \omega^2 \omega_0 - 3 \omega \omega_0^2 \big) \Big] \nonumber \\
    & \hspace{2cm} + 6 \lambda_E^2 \Big[e^{- \omega / \omega_0}\omega_0^3 \big(\gamma_E + ln \frac{\omega}{\omega_0} \big) - \omega_0^2\big\{(\omega_0 -\omega) Ei\Big(-\frac{\omega}{\omega_0} \Big) \big\}\Big]\Bigg\}\,.
\end{align}
This result is obtained with the following Grozin-Neubert constraints \cite{Grozin:1996pq}:
\begin{equation}
    \omega_0 = \lambda_B = \frac{2}{3} \Bar{\Lambda}\,, \hspace{1cm} 2 \Bar{\Lambda}^2 = 2 \lambda_E^2 + \lambda_H^2\,.
\end{equation}
\newpage
\textbf{Local duality A:}\\[1\baselineskip]
\begin{align}
    &\psi_5(\omega_1, \omega_2) = -\frac{15 \lambda_E^2}{6 \omega_0^5} \omega_2 \Big(\omega_0 - \frac{(\omega_1 + \omega_2)}{2} \Big)^2 \Theta(2\omega_0 - \omega_1 - \omega_2)\,,  \\
    & g_-(\omega) = -\frac{1}{384 \omega_0^5} \Theta(2\omega_0 - \omega) \Bigg\{\omega \Big((\omega - 2 \omega_0) (27 (\omega - 2 \omega_0)^2 \omega_0^2 + 10 \lambda_H^2 (7 \omega^2 - 12 \omega \omega_0 + 6 \omega_0^2)) \Big) \nonumber \\ 
     & \hspace{1cm} + 10 \lambda_E^2 \Big[(\omega - 2 \omega_0)^4 ln\Big(1 - \frac{\omega}{2\omega_0} \Big) - \omega^2(\omega^2 - 8 \omega \omega_0 + 24 \omega_0^2) ln \big(\frac{\omega}{\omega_0} \big) \nonumber \\
     & \hspace{1cm}+ \omega \Big(-12 \omega_0^3 + \frac{\omega^3}{2} (-15 + ln(4)) - 4 \omega^2\omega_0(-7+ln(4)) + 4 \omega \omega_0^2 (-5 +ln(64))\Big)\Big] \Bigg\}\,.
\end{align}
This result is obtained with the following Grozin-Neubert constraints \cite{Grozin:1996pq}:
\begin{equation}
    \omega_0 = \frac{3}{2}\lambda_B = \frac{4}{3} \Bar{\Lambda}\,, \hspace{1cm} 9 \omega_0^2 = 40(2 \lambda_E^2 + \lambda_H^2)\,.
\end{equation}

\textbf{Local duality B:}\\[1\baselineskip]
\begin{align}
    &\psi_5(\omega_1, \omega_2) = -\frac{35 \lambda_E^2}{64 \omega_0^7} \omega_2 \Big(2\omega_0 - \omega_1 - \omega_2 \Big)^4 \Theta(2\omega_0 - \omega_1 - \omega_2)\,,  \\
    & g_-(\omega) = \frac{1}{9126 \omega_0^7}\Theta(2\omega_0 - \omega) \Bigg\{\omega \Big[7 \lambda_E^2 (\omega - 2 \omega_0)(343 \omega^4 - 1798 \omega^3 \omega_0 + 3268 \omega^2 \omega_0^2 -2328 \omega \omega_0^3 -480 \omega_0^4)  \nonumber \\
    & \hspace{1cm} - 30 (\omega - 2 \omega_0)^3 (6 (\omega - 2 \omega_0)^2 \omega_0^2 + 7 \lambda_H^2 (11 \omega^2 - 12 \omega \omega_0 + 4 \omega_0^2)) \Big] \nonumber \\
    & \hspace{1cm} + 84 \lambda_E^2 \Big[ \omega^2 (\omega^4 - 12 \omega^3 \omega_0 + 60 \omega^2 \omega_0^2 - 160 \omega \omega_0^3 + 240 \omega_0^4) ln\Big(\frac{\omega}{2\omega_0 - \omega} \Big) \nonumber \\
    & \hspace{1cm} + 64 \omega_0^5 (-3 \omega + \omega_0) ln\Big(\frac{2 \omega_0}{2\omega_0 - \omega} \Big)\Big] \Bigg\}\,.
\end{align}
This result is obtained with the following Grozin-Neubert constraints \cite{Grozin:1996pq}:
\begin{equation}
    \omega_0 = \frac{5}{2}\lambda_B = 2 \Bar{\Lambda}\,, \hspace{1cm} 3 \omega_0^2 = 14 (2 \lambda_E^2 + \lambda_H^2)\,.
\end{equation}

\section{Light-meson LCDAs}
\label{appendix:lightmesonDAs}
For LCSRs with light-meson LCDAs, the starting point is a vacuum to light-meson correlation function \cite{Ball:2004rg, Ball:2004ye}:
\begin{equation}
    \Pi_\mu(q, p_B) = i \int d^4x e^{iq.x}\bra{M(k)}T J_\mu^{\text{weak}}(x) j_B^\dag(0) \ket{0}\,,
\end{equation}
where $J_\mu^{\text{weak}}$ is the relevant weak current, and $j_B$ is the interpolating field for the $B$ meson. \\
This time, the $B$ meson of momentum $p_B$ is off-shell while the light-meson $M$ is on-shell. Once again, the correlation function can be rewritten using analyticity and the unitary relation, and then further expressed as a sum of scalar functions times a Lorentz structure. Scalar relations are extracted in the form:
\begin{equation} \label{eq:Pi_p0_LM}
    \Pi_{F}(q^2,p_B^2) = Y_F \frac{F(q^2)}{m_B^2 - p_B^2} + \int_{s_{cont}}^\infty \frac{\rho_F(s)}{s-p_B^2}\,.
\end{equation}
On the LCOPE side, it takes the expression: 
\begin{equation}
    \Pi_F^{\text{LCOPE}}(q^2, p_B^2) = \frac{1}{\pi} \int_{m_b^2}^{\infty}ds \frac{\text{Im}\Pi_F^{\textrm{LCOPE}}(s)}{s-p_B^2}\,.
\end{equation}
The method presented in this article is applicable to the correlation function with light-meson LCDAs, and since similarly $m_B^2 < s_{cont}$ and $p_B^2<0$, we derive the analogous relations 
\begin{equation}\label{eq:lim_FF_LM}
    F(q^2) = \lim_{p\to\infty}  \frac{(m_B^2 - p_B^2)^{p+1}}{p! \, Y_F} \Pi_F^{(p)}(q^2,p_B^2)\,.
\end{equation}
And the corollary is:
\begin{equation}\label{eq:lim_m2_LM}
        m_B^2 = \lim_{p\to\infty}  \left[ \frac{p!}{(p-\ell)!}  \frac{\Pi_F^{(p-\ell)}}{\Pi_F^{(p)}} \right]^{1 / \ell} + p_B^2\,, \quad p>1 \,, \; p> \ell \geq 1\,.
\end{equation}

\section{Additional figures for $\widetilde{\Pi}_{F,LCOPE}^{(p)}$ and $\widetilde{m}_M^2$} \label{appendix:extrafigures}

In this section, we plot our numerical results for $\widetilde{\Pi}_{F,LCOPE}^{(p)}$ and $\widetilde{m}_M^2$ near the breakdown of the LCOPE for $f_+^{B \to \pi}$, $f_T^{B \to K, \pi}$, $V^{B\to \rho, K^*},A_1^{B\to \rho, K^*},T_1^{B\to \rho, K^*}$.

\begin{figure}[H]
    \centering
    \includegraphics[width=\textwidth]{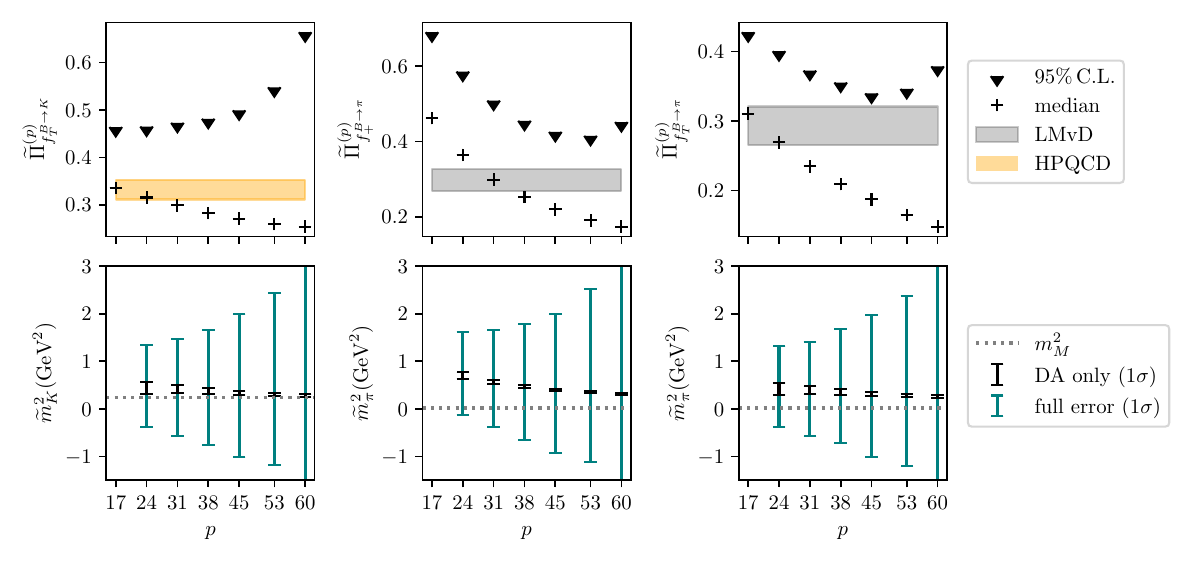}
    \caption{Results for $f_T^{B \to K}, \, f_+^{B \to \pi}, \, f_T^{B \to \pi} $ and the corresponding meson mass predictions at $k^2 = -20 \textrm{GeV}^2$. For comparison we also show the $1\sigma$ interval predicted in \cite{Parrott:2022rgu} and \cite{Leljak:2021vte}.}
    \label{fig:BtoP}
\end{figure}

\begin{figure}[H]
    \centering
    \includegraphics[width=\textwidth]{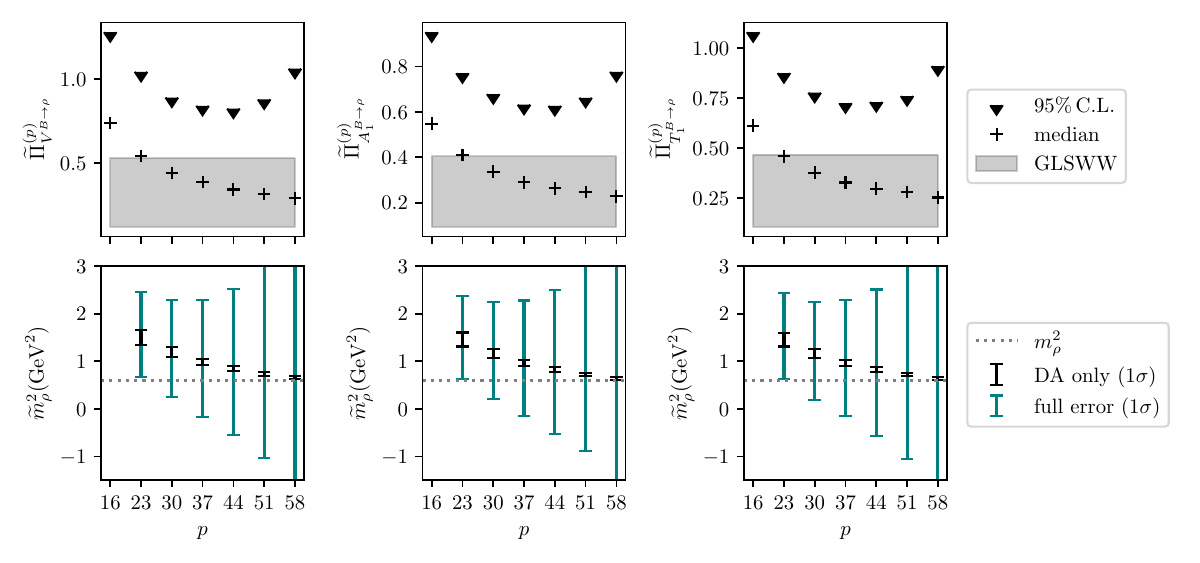}
    \caption{Results for $V^{B \to \rho}, \, A_1^{B \to \rho}, \, T_1^{B \to \rho} $ and the corresponding meson mass predictions at $k^2 = -20 \textrm{GeV}^2$. For comparison we also show the $1\sigma$ interval predicted in \cite{Gao:2019lta}.}
    \label{fig:Btotho}
\end{figure}

\begin{figure}[H]
    \centering
    \includegraphics[width=\textwidth]{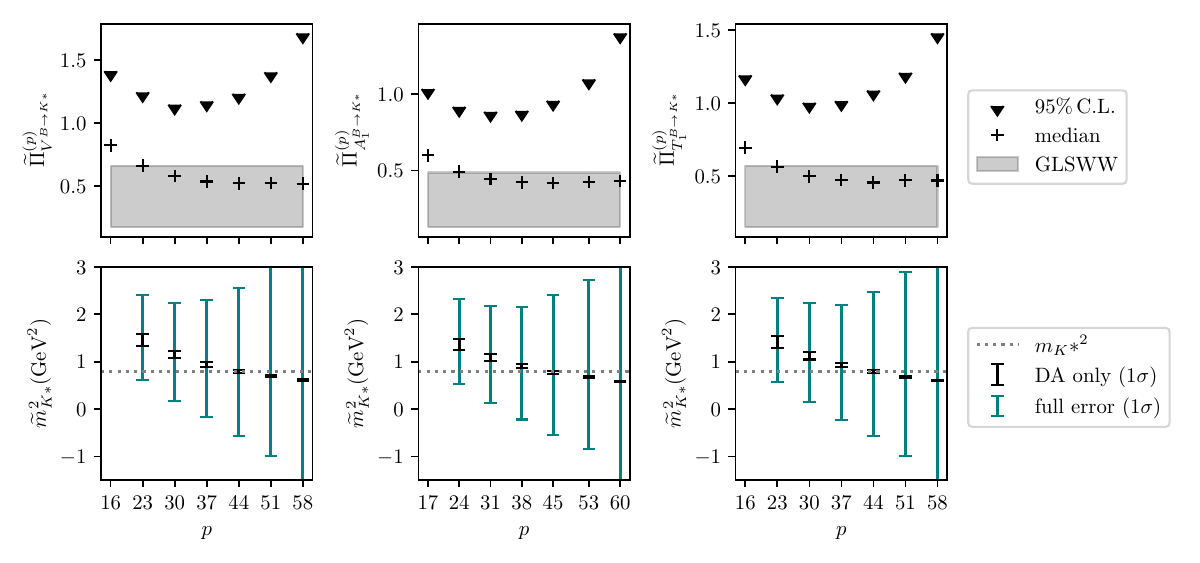}
    \caption{Results for $V^{B \to K^*}, \, A_1^{B \to K^*}, \, T_1^{B \to K^*} $ and the corresponding meson mass predictions at $k^2 = -20 \textrm{GeV}^2$. For comparison we also show the $1\sigma$ interval predicted in \cite{Gao:2019lta}.} 
    \label{fig:BtoKstar}
\end{figure}

\newpage
\addcontentsline{toc}{section}{References}
\bibliographystyle{JHEP}
\bibliography{bibliv3}

\providecommand{\href}[2]{#2}\begingroup\raggedright\begin{thebibliography}{10}

\bibitem{LHCb:2017avl}
{\scshape LHCb} collaboration, \emph{{Test of lepton universality with $B^{0}
  \rightarrow K^{*0}\ell^{+}\ell^{-}$ decays}},
  \href{https://doi.org/10.1007/JHEP08(2017)055}{\emph{JHEP} {\bfseries 08}
  (2017) 055} [\href{https://arxiv.org/abs/1705.05802}{{\ttfamily
  1705.05802}}].

\bibitem{LHCb:2021trn}
{\scshape LHCb} collaboration, \emph{{Test of lepton universality in
  beauty-quark decays}},
  \href{https://doi.org/10.1038/s41567-021-01478-8}{\emph{Nature Phys.}
  {\bfseries 18} (2022) 277}
  [\href{https://arxiv.org/abs/2103.11769}{{\ttfamily 2103.11769}}].

\bibitem{LHCb:2022vje}
{\scshape LHCb} collaboration, \emph{{Measurement of lepton universality
  parameters in $B^+\to K^+\ell^+\ell^-$ and $B^0\to K^{*0}\ell^+\ell^-$
  decays}}, \href{https://doi.org/10.1103/PhysRevD.108.032002}{\emph{Phys. Rev.
  D} {\bfseries 108} (2023) 032002}
  [\href{https://arxiv.org/abs/2212.09153}{{\ttfamily 2212.09153}}].

\bibitem{LHCb:2013ghj}
{\scshape LHCb} collaboration, \emph{{Measurement of Form-Factor-Independent
  Observables in the Decay $B^{0} \to K^{*0} \mu^+ \mu^-$}},
  \href{https://doi.org/10.1103/PhysRevLett.111.191801}{\emph{Phys. Rev. Lett.}
  {\bfseries 111} (2013) 191801}
  [\href{https://arxiv.org/abs/1308.1707}{{\ttfamily 1308.1707}}].

\bibitem{LHCb:2014cxe}
{\scshape LHCb} collaboration, \emph{{Differential branching fractions and
  isospin asymmetries of $B \to K^{(*)} \mu^+ \mu^-$ decays}},
  \href{https://doi.org/10.1007/JHEP06(2014)133}{\emph{JHEP} {\bfseries 06}
  (2014) 133} [\href{https://arxiv.org/abs/1403.8044}{{\ttfamily 1403.8044}}].

\bibitem{LHCb:2015tgy}
{\scshape LHCb} collaboration, \emph{{Differential branching fraction and
  angular analysis of $\Lambda^{0}_{b} \rightarrow \Lambda \mu^+\mu^-$
  decays}}, \href{https://doi.org/10.1007/JHEP06(2015)115}{\emph{JHEP}
  {\bfseries 06} (2015) 115}
  [\href{https://arxiv.org/abs/1503.07138}{{\ttfamily 1503.07138}}].

\bibitem{LHCb:2015wdu}
{\scshape LHCb} collaboration, \emph{{Angular analysis and differential
  branching fraction of the decay $B^0_s\to\phi\mu^+\mu^-$}},
  \href{https://doi.org/10.1007/JHEP09(2015)179}{\emph{JHEP} {\bfseries 09}
  (2015) 179} [\href{https://arxiv.org/abs/1506.08777}{{\ttfamily
  1506.08777}}].

\bibitem{LHCb:2015svh}
{\scshape LHCb} collaboration, \emph{{Angular analysis of the $B^{0} \to K^{*0}
  \mu^{+} \mu^{-}$ decay using 3 fb$^{-1}$ of integrated luminosity}},
  \href{https://doi.org/10.1007/JHEP02(2016)104}{\emph{JHEP} {\bfseries 02}
  (2016) 104} [\href{https://arxiv.org/abs/1512.04442}{{\ttfamily
  1512.04442}}].

\bibitem{LHCb:2020lmf}
{\scshape LHCb} collaboration, \emph{{Measurement of $CP$-Averaged Observables
  in the $B^{0}\rightarrow K^{*0}\mu^{+}\mu^{-}$ Decay}},
  \href{https://doi.org/10.1103/PhysRevLett.125.011802}{\emph{Phys. Rev. Lett.}
  {\bfseries 125} (2020) 011802}
  [\href{https://arxiv.org/abs/2003.04831}{{\ttfamily 2003.04831}}].

\bibitem{LHCb:2020gog}
{\scshape LHCb} collaboration, \emph{{Angular Analysis of the $B^{+}\rightarrow
  K^{\ast+}\mu^{+}\mu^{-}$ Decay}},
  \href{https://doi.org/10.1103/PhysRevLett.126.161802}{\emph{Phys. Rev. Lett.}
  {\bfseries 126} (2021) 161802}
  [\href{https://arxiv.org/abs/2012.13241}{{\ttfamily 2012.13241}}].

\bibitem{LHCb:2021xxq}
{\scshape LHCb} collaboration, \emph{{Angular analysis of the rare decay
  $B^0_s\to\phi\mu^+\mu^-$}},
  \href{https://doi.org/10.1007/JHEP11(2021)043}{\emph{JHEP} {\bfseries 11}
  (2021) 043} [\href{https://arxiv.org/abs/2107.13428}{{\ttfamily
  2107.13428}}].

\bibitem{LHCb:2021zwz}
{\scshape LHCb} collaboration, \emph{{Branching Fraction Measurements of the
  Rare $B^0_s\rightarrow\phi\mu^+\mu^-$ and $B^0_s\rightarrow
  f_2^\prime(1525)\mu^+\mu^-$- Decays}},
  \href{https://doi.org/10.1103/PhysRevLett.127.151801}{\emph{Phys. Rev. Lett.}
  {\bfseries 127} (2021) 151801}
  [\href{https://arxiv.org/abs/2105.14007}{{\ttfamily 2105.14007}}].

\bibitem{CMS:2024syx}
{\scshape CMS} collaboration, \emph{{Test of lepton flavor universality in
  B$^{\pm}$$\to$ K$^{\pm}\mu^+\mu^-$ and B$^{\pm}$$\to$ K$^{\pm}$e$^+$e$^-$
  decays in proton-proton collisions at $\sqrt{s}$ = 13 TeV}},
  \href{https://doi.org/10.1088/1361-6633/ad4e65}{\emph{Rept. Prog. Phys.}
  {\bfseries 87} (2024) 077802}
  [\href{https://arxiv.org/abs/2401.07090}{{\ttfamily 2401.07090}}].

\bibitem{CMS-PAS-BPH-21-002}
{\scshape CMS} collaboration, \emph{{Angular analysis of the $B^0 \to
  K^{*0}(892) \mu^+ \mu^-$ decay at $\sqrt{s}$ = 13 TeV}},  Tech. Rep.
  \href{https://cds.cern.ch/record/2899589}{}, CERN, Geneva (2024).

\bibitem{Bobeth:2017vxj}
C.~Bobeth, M.~Chrzaszcz, D.~van Dyk and J.~Virto, \emph{{Long-distance effects
  in $B\rightarrow K^*\ell \ell $ from analyticity}},
  \href{https://doi.org/10.1140/epjc/s10052-018-5918-6}{\emph{Eur. Phys. J. C}
  {\bfseries 78} (2018) 451}
  [\href{https://arxiv.org/abs/1707.07305}{{\ttfamily 1707.07305}}].

\bibitem{Gubernari:2020eft}
N.~Gubernari, D.~van Dyk and J.~Virto, \emph{{Non-local matrix elements in
  $B_{(s)}\to \{K^{(*)},\phi\}\ell^+\ell^-$}},
  \href{https://doi.org/10.1007/JHEP02(2021)088}{\emph{JHEP} {\bfseries 02}
  (2021) 088} [\href{https://arxiv.org/abs/2011.09813}{{\ttfamily
  2011.09813}}].

\bibitem{Jager:2012uw}
S.~J\"ager and J.~Martin~Camalich, \emph{{On $B \to V \ell \ell$ at small
  dilepton invariant mass, power corrections, and new physics}},
  \href{https://doi.org/10.1007/JHEP05(2013)043}{\emph{JHEP} {\bfseries 05}
  (2013) 043} [\href{https://arxiv.org/abs/1212.2263}{{\ttfamily 1212.2263}}].

\bibitem{Jager:2014rwa}
S.~J\"ager and J.~Martin~Camalich, \emph{{Reassessing the discovery potential
  of the $B \to K^{*} \ell^+\ell^-$ decays in the large-recoil region: SM
  challenges and BSM opportunities}},
  \href{https://doi.org/10.1103/PhysRevD.93.014028}{\emph{Phys. Rev. D}
  {\bfseries 93} (2016) 014028}
  [\href{https://arxiv.org/abs/1412.3183}{{\ttfamily 1412.3183}}].

\bibitem{Ciuchini:2015qxb}
M.~Ciuchini, M.~Fedele, E.~Franco, S.~Mishima, A.~Paul, L.~Silvestrini et~al.,
  \emph{{$B\to K^* \ell^+ \ell^-$ decays at large recoil in the Standard Model:
  a theoretical reappraisal}},
  \href{https://doi.org/10.1007/JHEP06(2016)116}{\emph{JHEP} {\bfseries 06}
  (2016) 116} [\href{https://arxiv.org/abs/1512.07157}{{\ttfamily
  1512.07157}}].

\bibitem{Capdevila:2017ert}
B.~Capdevila, S.~Descotes-Genon, L.~Hofer and J.~Matias, \emph{{Hadronic
  uncertainties in $B \to K^* \mu^+ \mu^-$: a state-of-the-art analysis}},
  \href{https://doi.org/10.1007/JHEP04(2017)016}{\emph{JHEP} {\bfseries 04}
  (2017) 016} [\href{https://arxiv.org/abs/1701.08672}{{\ttfamily
  1701.08672}}].

\bibitem{Chobanova:2017ghn}
V.G.~Chobanova, T.~Hurth, F.~Mahmoudi, D.~Martinez~Santos and S.~Neshatpour,
  \emph{{Large hadronic power corrections or new physics in the rare decay
  $B\to K^{*}\mu^{+}\mu^{-}$?}},
  \href{https://doi.org/10.1007/JHEP07(2017)025}{\emph{JHEP} {\bfseries 07}
  (2017) 025} [\href{https://arxiv.org/abs/1702.02234}{{\ttfamily
  1702.02234}}].

\bibitem{Arbey:2018ics}
A.~Arbey, T.~Hurth, F.~Mahmoudi and S.~Neshatpour, \emph{{Hadronic and New
  Physics Contributions to $b \to s$ Transitions}},
  \href{https://doi.org/10.1103/PhysRevD.98.095027}{\emph{Phys. Rev. D}
  {\bfseries 98} (2018) 095027}
  [\href{https://arxiv.org/abs/1806.02791}{{\ttfamily 1806.02791}}].

\bibitem{Ciuchini:2018anp}
M.~Ciuchini, A.M.~Coutinho, M.~Fedele, E.~Franco, A.~Paul, L.~Silvestrini
  et~al., \emph{{Hadronic uncertainties in semileptonic $B\to K^*\mu^+\mu^-$
  decays}}, \href{https://doi.org/10.22323/1.326.0044}{\emph{PoS} {\bfseries
  BEAUTY2018} (2018) 044} [\href{https://arxiv.org/abs/1809.03789}{{\ttfamily
  1809.03789}}].

\bibitem{Hurth:2020rzx}
T.~Hurth, F.~Mahmoudi and S.~Neshatpour, \emph{{Implications of the new LHCb
  angular analysis of $B \to K^* \mu^+ \mu^-$ : Hadronic effects or new
  physics?}}, \href{https://doi.org/10.1103/PhysRevD.102.055001}{\emph{Phys.
  Rev. D} {\bfseries 102} (2020) 055001}
  [\href{https://arxiv.org/abs/2006.04213}{{\ttfamily 2006.04213}}].

\bibitem{Bordone:2024hui}
M.~Bordone, G.~isidori, S.~M\"achler and A.~Tinari, \emph{{Short- vs.
  long-distance physics in $B\rightarrow K^{(*)} \ell ^+\ell ^-$: a data-driven
  analysis}}, \href{https://doi.org/10.1140/epjc/s10052-024-12869-5}{\emph{Eur.
  Phys. J. C} {\bfseries 84} (2024) 547}
  [\href{https://arxiv.org/abs/2401.18007}{{\ttfamily 2401.18007}}].

\bibitem{Mahmoudi:2024zna}
F.~Mahmoudi and Y.~Monceaux, \emph{{Overview of $B \to K^{(*)}\ell \ell$
  Theoretical Calculations and Uncertainties}},
  \href{https://doi.org/10.3390/sym1608100}{\emph{Symmetry} {\bfseries 16}
  (2024) 1006} [\href{https://arxiv.org/abs/2408.03235}{{\ttfamily
  2408.03235}}].

\bibitem{Parrott:2022rgu}
{\scshape HPQCD} collaboration, \emph{{B\textrightarrow{}K and
  D\textrightarrow{}K form factors from fully relativistic lattice QCD}},
  \href{https://doi.org/10.1103/PhysRevD.107.014510}{\emph{Phys. Rev. D}
  {\bfseries 107} (2023) 014510}
  [\href{https://arxiv.org/abs/2207.12468}{{\ttfamily 2207.12468}}].

\bibitem{Khodjamirian:1997lay}
A.~Khodjamirian and R.~Ruckl, \emph{{QCD sum rules for exclusive decays of
  heavy mesons}}, \href{https://doi.org/10.1142/9789812812667_0005}{\emph{Adv.
  Ser. Direct. High Energy Phys.} {\bfseries 15} (1998) 345}
  [\href{https://arxiv.org/abs/hep-ph/9801443}{{\ttfamily hep-ph/9801443}}].

\bibitem{Braun:1999dp}
V.M.~Braun, \emph{{QCD sum rules for heavy flavors}},
  \href{https://doi.org/10.22323/1.003.0006}{\emph{PoS} {\bfseries hf8} (1999)
  006} [\href{https://arxiv.org/abs/hep-ph/9911206}{{\ttfamily
  hep-ph/9911206}}].

\bibitem{Ball:2004rg}
P.~Ball and R.~Zwicky, \emph{{$B_{d,s} \to \rho, \omega, K^*, \phi$ decay
  form-factors from light-cone sum rules revisited}},
  \href{https://doi.org/10.1103/PhysRevD.71.014029}{\emph{Phys. Rev. D}
  {\bfseries 71} (2005) 014029}
  [\href{https://arxiv.org/abs/hep-ph/0412079}{{\ttfamily hep-ph/0412079}}].

\bibitem{Ball:2004ye}
P.~Ball and R.~Zwicky, \emph{{New results on $B \to \pi, K, \eta$ decay
  formfactors from light-cone sum rules}},
  \href{https://doi.org/10.1103/PhysRevD.71.014015}{\emph{Phys. Rev. D}
  {\bfseries 71} (2005) 014015}
  [\href{https://arxiv.org/abs/hep-ph/0406232}{{\ttfamily hep-ph/0406232}}].

\bibitem{Khodjamirian:2005ea}
A.~Khodjamirian, T.~Mannel and N.~Offen, \emph{{B-meson distribution amplitude
  from the $B \to \pi$ form-factor}},
  \href{https://doi.org/10.1016/j.physletb.2005.06.021}{\emph{Phys. Lett. B}
  {\bfseries 620} (2005) 52}
  [\href{https://arxiv.org/abs/hep-ph/0504091}{{\ttfamily hep-ph/0504091}}].

\bibitem{Khodjamirian:2006st}
A.~Khodjamirian, T.~Mannel and N.~Offen, \emph{{Form-factors from light-cone
  sum rules with B-meson distribution amplitudes}},
  \href{https://doi.org/10.1103/PhysRevD.75.054013}{\emph{Phys. Rev. D}
  {\bfseries 75} (2007) 054013}
  [\href{https://arxiv.org/abs/hep-ph/0611193}{{\ttfamily hep-ph/0611193}}].

\bibitem{Duplancic:2008ix}
G.~Duplancic, A.~Khodjamirian, T.~Mannel, B.~Melic and N.~Offen,
  \emph{{Light-cone sum rules for $B \to \pi$ form factors revisited}},
  \href{https://doi.org/10.1088/1126-6708/2008/04/014}{\emph{JHEP} {\bfseries
  04} (2008) 014} [\href{https://arxiv.org/abs/0801.1796}{{\ttfamily
  0801.1796}}].

\bibitem{Duplancic:2008tk}
G.~Duplancic and B.~Melic, \emph{{B, B(s) $\to$ K form factors: An Update of
  light-cone sum rule results}},
  \href{https://doi.org/10.1103/PhysRevD.78.054015}{\emph{Phys. Rev. D}
  {\bfseries 78} (2008) 054015}
  [\href{https://arxiv.org/abs/0805.4170}{{\ttfamily 0805.4170}}].

\bibitem{Khodjamirian:2010vf}
A.~Khodjamirian, T.~Mannel, A.A.~Pivovarov and Y.M.~Wang, \emph{{Charm-loop
  effect in $B \to K^{(*)} \ell^{+} \ell^{-}$ and $B\to K^*\gamma$}},
  \href{https://doi.org/10.1007/JHEP09(2010)089}{\emph{JHEP} {\bfseries 09}
  (2010) 089} [\href{https://arxiv.org/abs/1006.4945}{{\ttfamily 1006.4945}}].

\bibitem{Bharucha:2012wy}
A.~Bharucha, \emph{{Two-loop Corrections to the $B to \pi$ Form Factor from QCD
  Sum Rules on the Light-Cone and $|V_{ub}|$}},
  \href{https://doi.org/10.1007/JHEP05(2012)092}{\emph{JHEP} {\bfseries 05}
  (2012) 092} [\href{https://arxiv.org/abs/1203.1359}{{\ttfamily 1203.1359}}].

\bibitem{Wang:2015vgv}
Y.-M.~Wang and Y.-L.~Shen, \emph{{QCD corrections to B \textrightarrow{}
  \ensuremath{\pi} form factors from light-cone sum rules}},
  \href{https://doi.org/10.1016/j.nuclphysb.2015.07.016}{\emph{Nucl. Phys. B}
  {\bfseries 898} (2015) 563}
  [\href{https://arxiv.org/abs/1506.00667}{{\ttfamily 1506.00667}}].

\bibitem{Rusov:2017chr}
A.V.~Rusov, \emph{{Higher-twist effects in light-cone sum rule for the
  $B\rightarrow \pi $ form factor}},
  \href{https://doi.org/10.1140/epjc/s10052-017-5000-9}{\emph{Eur. Phys. J. C}
  {\bfseries 77} (2017) 442}
  [\href{https://arxiv.org/abs/1705.01929}{{\ttfamily 1705.01929}}].

\bibitem{Khodjamirian:2017fxg}
A.~Khodjamirian and A.V.~Rusov, \emph{{$B_{s}\to K \ell \nu_\ell$ and $B_{(s)}
  \to \pi (K) \ell^+\ell^-$ decays at large recoil and CKM matrix elements}},
  \href{https://doi.org/10.1007/JHEP08(2017)112}{\emph{JHEP} {\bfseries 08}
  (2017) 112} [\href{https://arxiv.org/abs/1703.04765}{{\ttfamily
  1703.04765}}].

\bibitem{Lu:2018cfc}
C.-D.~L\"u, Y.-L.~Shen, Y.-M.~Wang and Y.-B.~Wei, \emph{{QCD calculations of $B
  \to \pi, K$ form factors with higher-twist corrections}},
  \href{https://doi.org/10.1007/JHEP01(2019)024}{\emph{JHEP} {\bfseries 01}
  (2019) 024} [\href{https://arxiv.org/abs/1810.00819}{{\ttfamily
  1810.00819}}].

\bibitem{Cui:2022zwm}
B.-Y.~Cui, Y.-K.~Huang, Y.-L.~Shen, C.~Wang and Y.-M.~Wang, \emph{{Precision
  calculations of B$_{d,s}$ \textrightarrow{} \ensuremath{\pi}, K decay form
  factors in soft-collinear effective theory}},
  \href{https://doi.org/10.1007/JHEP03(2023)140}{\emph{JHEP} {\bfseries 03}
  (2023) 140} [\href{https://arxiv.org/abs/2212.11624}{{\ttfamily
  2212.11624}}].

\bibitem{Monceaux:2023byy}
Y.~Monceaux, A.~Carvunis and F.~Mahmoudi, \emph{{LCSR predictions for $b \to s$
  hadronic form factors}},
  \href{https://doi.org/10.22323/1.445.0060}{\emph{PoS} {\bfseries FPCP2023}
  (2023) 060}.

\bibitem{Gubernari:2023rfu}
N.~Gubernari, A.~Khodjamirian, R.~Mandal and T.~Mannel, \emph{{$ B\to
  {D}_0^{\ast } $ and $ {B}_s\to {D}_{s0}^{\ast } $ form factors from QCD
  light-cone sum rules}},
  \href{https://doi.org/10.1007/JHEP12(2023)015}{\emph{JHEP} {\bfseries 12}
  (2023) 015} [\href{https://arxiv.org/abs/2309.10165}{{\ttfamily
  2309.10165}}].

\bibitem{Khodjamirian:2023wol}
A.~Khodjamirian, B.~Meli\'c and Y.-M.~Wang, \emph{{A guide to the QCD
  light-cone sum rules for b-quark decays}},
  \href{https://doi.org/10.1140/epjs/s11734-023-01046-6}{\emph{Eur. Phys. J.
  ST} {\bfseries 233} (2024) 271}
  [\href{https://arxiv.org/abs/2311.08700}{{\ttfamily 2311.08700}}].

\bibitem{Grozin:1996pq}
A.G.~Grozin and M.~Neubert, \emph{{Asymptotics of heavy meson form-factors}},
  \href{https://doi.org/10.1103/PhysRevD.55.272}{\emph{Phys. Rev. D} {\bfseries
  55} (1997) 272} [\href{https://arxiv.org/abs/hep-ph/9607366}{{\ttfamily
  hep-ph/9607366}}].

\bibitem{Beneke:2000wa}
M.~Beneke and T.~Feldmann, \emph{{Symmetry breaking corrections to heavy to
  light B meson form-factors at large recoil}},
  \href{https://doi.org/10.1016/S0550-3213(00)00585-X}{\emph{Nucl. Phys. B}
  {\bfseries 592} (2001) 3}
  [\href{https://arxiv.org/abs/hep-ph/0008255}{{\ttfamily hep-ph/0008255}}].

\bibitem{DeFazio:2005dx}
F.~De~Fazio, T.~Feldmann and T.~Hurth, \emph{{Light-cone sum rules in
  soft-collinear effective theory}},
  \href{https://doi.org/10.1016/j.nuclphysb.2008.03.022}{\emph{Nucl. Phys. B}
  {\bfseries 733} (2006) 1}
  [\href{https://arxiv.org/abs/hep-ph/0504088}{{\ttfamily hep-ph/0504088}}].

\bibitem{Kawamura:2001bp}
H.~Kawamura, J.~Kodaira, C.-F.~Qiao and K.~Tanaka, \emph{{B meson light cone
  distribution amplitudes and heavy quark symmetry}},
  \href{https://doi.org/10.1142/S0217751X03014848}{\emph{Int. J. Mod. Phys. A}
  {\bfseries 18} (2003) 1433}
  [\href{https://arxiv.org/abs/hep-ph/0112146}{{\ttfamily hep-ph/0112146}}].

\bibitem{Kawamura:2002mg}
H.~Kawamura, J.~Kodaira, C.-F.~Qiao and K.~Tanaka, \emph{{B meson light cone
  wavefunctions in the heavy quark limit}},
  \href{https://doi.org/10.1016/S0920-5632(03)80182-2}{\emph{Nucl. Phys. B
  Proc. Suppl.} {\bfseries 116} (2003) 269}
  [\href{https://arxiv.org/abs/hep-ph/0211270}{{\ttfamily hep-ph/0211270}}].

\bibitem{Lange:2003ff}
B.O.~Lange and M.~Neubert, \emph{{Renormalization group evolution of the B
  meson light cone distribution amplitude}},
  \href{https://doi.org/10.1103/PhysRevLett.91.102001}{\emph{Phys. Rev. Lett.}
  {\bfseries 91} (2003) 102001}
  [\href{https://arxiv.org/abs/hep-ph/0303082}{{\ttfamily hep-ph/0303082}}].

\bibitem{DeFazio:2007hw}
F.~De~Fazio, T.~Feldmann and T.~Hurth, \emph{{SCET sum rules for $B \to P$ and
  $B \to V$ transition form factors}},
  \href{https://doi.org/10.1088/1126-6708/2008/02/031}{\emph{JHEP} {\bfseries
  02} (2008) 031} [\href{https://arxiv.org/abs/0711.3999}{{\ttfamily
  0711.3999}}].

\bibitem{Kawamura:2010tj}
H.~Kawamura and K.~Tanaka, \emph{{Evolution equation for the B-meson
  distribution amplitude in the heavy-quark effective theory in coordinate
  space}}, \href{https://doi.org/10.1103/PhysRevD.81.114009}{\emph{Phys. Rev.
  D} {\bfseries 81} (2010) 114009}
  [\href{https://arxiv.org/abs/1002.1177}{{\ttfamily 1002.1177}}].

\bibitem{Bell:2013tfa}
G.~Bell, T.~Feldmann, Y.-M.~Wang and M.W.Y.~Yip, \emph{{Light-Cone Distribution
  Amplitudes for Heavy-Quark Hadrons}},
  \href{https://doi.org/10.1007/JHEP11(2013)191}{\emph{JHEP} {\bfseries 11}
  (2013) 191} [\href{https://arxiv.org/abs/1308.6114}{{\ttfamily 1308.6114}}].

\bibitem{Feldmann:2014ika}
T.~Feldmann, B.O.~Lange and Y.-M.~Wang, \emph{{B -meson light-cone distribution
  amplitude: Perturbative constraints and asymptotic behavior in dual space}},
  \href{https://doi.org/10.1103/PhysRevD.89.114001}{\emph{Phys. Rev. D}
  {\bfseries 89} (2014) 114001}
  [\href{https://arxiv.org/abs/1404.1343}{{\ttfamily 1404.1343}}].

\bibitem{Braun:2014owa}
V.M.~Braun and A.N.~Manashov, \emph{{Conformal symmetry of the Lange-Neubert
  evolution equation}},
  \href{https://doi.org/10.1016/j.physletb.2014.02.051}{\emph{Phys. Lett. B}
  {\bfseries 731} (2014) 316}
  [\href{https://arxiv.org/abs/1402.5822}{{\ttfamily 1402.5822}}].

\bibitem{Braun:2015pha}
V.M.~Braun, A.N.~Manashov and N.~Offen, \emph{{Evolution equation for the
  higher-twist B-meson distribution amplitude}},
  \href{https://doi.org/10.1103/PhysRevD.92.074044}{\emph{Phys. Rev. D}
  {\bfseries 92} (2015) 074044}
  [\href{https://arxiv.org/abs/1507.03445}{{\ttfamily 1507.03445}}].

\bibitem{Braun:2017liq}
V.M.~Braun, Y.~Ji and A.N.~Manashov, \emph{{Higher-twist B-meson Distribution
  Amplitudes in HQET}},
  \href{https://doi.org/10.1007/JHEP05(2017)022}{\emph{JHEP} {\bfseries 05}
  (2017) 022} [\href{https://arxiv.org/abs/1703.02446}{{\ttfamily
  1703.02446}}].

\bibitem{Braun:2019wyx}
V.M.~Braun, Y.~Ji and A.N.~Manashov, \emph{{Two-loop evolution equation for the
  B-meson distribution amplitude}},
  \href{https://doi.org/10.3204/PUBDB-2019-02451}{\emph{Phys. Rev. D}
  {\bfseries 100} (2019) 014023}
  [\href{https://arxiv.org/abs/1905.04498}{{\ttfamily 1905.04498}}].

\bibitem{Feldmann:2022uok}
T.~Feldmann, P.~L\"ughausen and D.~van Dyk, \emph{{Systematic parametrization
  of the leading B-meson light-cone distribution amplitude}},
  \href{https://doi.org/10.1007/JHEP10(2022)162}{\emph{JHEP} {\bfseries 10}
  (2022) 162} [\href{https://arxiv.org/abs/2203.15679}{{\ttfamily
  2203.15679}}].

\bibitem{Gubernari:2018wyi}
N.~Gubernari, A.~Kokulu and D.~van Dyk, \emph{{$B\to P$ and $B\to V$ Form
  Factors from $B$-Meson Light-Cone Sum Rules beyond Leading Twist}},
  \href{https://doi.org/10.1007/JHEP01(2019)150}{\emph{JHEP} {\bfseries 01}
  (2019) 150} [\href{https://arxiv.org/abs/1811.00983}{{\ttfamily
  1811.00983}}].

\bibitem{Khodjamirian:2020btr}
A.~Khodjamirian, \emph{{Hadron Form Factors}: {From Basic Phenomenology to QCD
  Sum Rules}}, CRC Press, Taylor \& Francis Group, Boca Raton, FL, USA (2020).

\bibitem{Colangelo:2000dp}
P.~Colangelo and A.~Khodjamirian, \emph{{QCD sum rules, a modern perspective}},
   in \emph{{At the frontier of particle physics. Handbook of QCD. Vol. 1-3}},
  M.~Shifman and B.~Ioffe, eds., (Singapore), pp.~1495--1576, World Scientific
  (2000) [\href{https://arxiv.org/abs/hep-ph/0010175}{{\ttfamily
  hep-ph/0010175}}].

\bibitem{Bharucha:2015bzk}
A.~Bharucha, D.M.~Straub and R.~Zwicky, \emph{{$B\to V\ell^+\ell^-$ in the
  Standard Model from light-cone sum rules}},
  \href{https://doi.org/10.1007/JHEP08(2016)098}{\emph{JHEP} {\bfseries 08}
  (2016) 098} [\href{https://arxiv.org/abs/1503.05534}{{\ttfamily
  1503.05534}}].

\bibitem{Braun:1989iv}
V.M.~Braun and I.E.~Filyanov, \emph{{Conformal Invariance and Pion Wave
  Functions of Nonleading Twist}},
  \href{https://doi.org/10.1007/BF01554472}{\emph{Z. Phys. C} {\bfseries 48}
  (1990) 239}.

\bibitem{Feldmann:2023aml}
T.~Feldmann, P.~L\"ughausen and N.~Seitz, \emph{{Strange-quark mass effects in
  the $B_{s}$ meson\textquoteright{}s light-cone distribution amplitude}},
  \href{https://doi.org/10.1007/JHEP08(2023)075}{\emph{JHEP} {\bfseries 08}
  (2023) 075} [\href{https://arxiv.org/abs/2306.14686}{{\ttfamily
  2306.14686}}].

\bibitem{Braun:2003wx}
V.M.~Braun, D.Y.~Ivanov and G.P.~Korchemsky, \emph{{The B meson distribution
  amplitude in QCD}},
  \href{https://doi.org/10.1103/PhysRevD.69.034014}{\emph{Phys. Rev. D}
  {\bfseries 69} (2004) 034014}
  [\href{https://arxiv.org/abs/hep-ph/0309330}{{\ttfamily hep-ph/0309330}}].

\bibitem{Khodjamirian:2020hob}
A.~Khodjamirian, R.~Mandal and T.~Mannel, \emph{{Inverse moment of the
  B$_{s}$-meson distribution amplitude from QCD sum rule}},
  \href{https://doi.org/10.1007/JHEP10(2020)043}{\emph{JHEP} {\bfseries 10}
  (2020) 043} [\href{https://arxiv.org/abs/2008.03935}{{\ttfamily
  2008.03935}}].

\bibitem{Nishikawa:2011qk}
T.~Nishikawa and K.~Tanaka, \emph{{QCD Sum Rules for Quark-Gluon Three-Body
  Components in the B Meson}},
  \href{https://doi.org/10.1016/j.nuclphysb.2013.12.007}{\emph{Nucl. Phys. B}
  {\bfseries 879} (2014) 110}
  [\href{https://arxiv.org/abs/1109.6786}{{\ttfamily 1109.6786}}].

\bibitem{Rahimi:2020zzo}
M.~Rahimi and M.~Wald, \emph{{QCD sum rules for parameters of the B-meson
  distribution amplitudes}},
  \href{https://doi.org/10.1103/PhysRevD.104.016027}{\emph{Phys. Rev. D}
  {\bfseries 104} (2021) 016027}
  [\href{https://arxiv.org/abs/2012.12165}{{\ttfamily 2012.12165}}].

\bibitem{Shifman:1978bx}
M.A.~Shifman, A.I.~Vainshtein and V.I.~Zakharov, \emph{{QCD and Resonance
  Physics. Theoretical Foundations}},
  \href{https://doi.org/10.1016/0550-3213(79)90022-1}{\emph{Nucl. Phys. B}
  {\bfseries 147} (1979) 385}.

\bibitem{Braun:1994ij}
V.M.~Braun and I.E.~Halperin, \emph{{Soft contribution to the pion form-factor
  from light cone QCD sum rules}},
  \href{https://doi.org/10.1016/0370-2693(94)91505-9}{\emph{Phys. Lett. B}
  {\bfseries 328} (1994) 457}
  [\href{https://arxiv.org/abs/hep-ph/9402270}{{\ttfamily hep-ph/9402270}}].

\bibitem{Khodjamirian:1997tk}
A.~Khodjamirian, \emph{{Form-factors of $\gamma^* \rho \to \pi$ and $\gamma^*
  \gamma \to \pi^0$ transitions and light cone sum rules}},
  \href{https://doi.org/10.1007/s100520050357}{\emph{Eur. Phys. J. C}
  {\bfseries 6} (1999) 477}
  [\href{https://arxiv.org/abs/hep-ph/9712451}{{\ttfamily hep-ph/9712451}}].

\bibitem{Braun:1999uj}
V.M.~Braun, A.~Khodjamirian and M.~Maul, \emph{{Pion form-factor in QCD at
  intermediate momentum transfers}},
  \href{https://doi.org/10.1103/PhysRevD.61.073004}{\emph{Phys. Rev. D}
  {\bfseries 61} (2000) 073004}
  [\href{https://arxiv.org/abs/hep-ph/9907495}{{\ttfamily hep-ph/9907495}}].

\bibitem{Descotes-Genon:2019bud}
S.~Descotes-Genon, A.~Khodjamirian and J.~Virto, \emph{{Light-cone sum rules
  for $B\to K\pi$ form factors and applications to rare decays}},
  \href{https://doi.org/10.1007/JHEP12(2019)083}{\emph{JHEP} {\bfseries 12}
  (2019) 083} [\href{https://arxiv.org/abs/1908.02267}{{\ttfamily
  1908.02267}}].

\bibitem{Descotes-Genon:2023ukb}
S.~Descotes-Genon, A.~Khodjamirian, J.~Virto and K.K.~Vos, \emph{{Light-Cone
  Sum Rules for $S$-wave $B\to K\pi$ Form Factors}},
  \href{https://doi.org/10.1007/JHEP06(2023)034}{\emph{JHEP} {\bfseries 06}
  (2023) 034} [\href{https://arxiv.org/abs/2304.02973}{{\ttfamily
  2304.02973}}].

\bibitem{Gelhausen:2013wia}
P.~Gelhausen, A.~Khodjamirian, A.A.~Pivovarov and D.~Rosenthal, \emph{{Decay
  constants of heavy-light vector mesons from QCD sum rules}},
  \href{https://doi.org/10.1103/PhysRevD.88.014015}{\emph{Phys. Rev. D}
  {\bfseries 88} (2013) 014015}
  [\href{https://arxiv.org/abs/1305.5432}{{\ttfamily 1305.5432}}].

\bibitem{Khodjamirian:2008xt}
A.~Khodjamirian, \emph{{Upper bounds on f(D) and f(D(s)) from two-point
  correlation function in QCD}},
  \href{https://doi.org/10.1103/PhysRevD.79.031503}{\emph{Phys. Rev. D}
  {\bfseries 79} (2009) 031503}
  [\href{https://arxiv.org/abs/0812.3747}{{\ttfamily 0812.3747}}].

\bibitem{Herren:2017osy}
F.~Herren and M.~Steinhauser, \emph{{Version 3 of RunDec and CRunDec}},
  \href{https://doi.org/10.1016/j.cpc.2017.11.014}{\emph{Comput. Phys. Commun.}
  {\bfseries 224} (2018) 333}
  [\href{https://arxiv.org/abs/1703.03751}{{\ttfamily 1703.03751}}].

\bibitem{FlavourLatticeAveragingGroupFLAG:2021npn}
{\scshape Flavour Lattice Averaging Group (FLAG)} collaboration, \emph{{FLAG
  Review 2021}},
  \href{https://doi.org/10.1140/epjc/s10052-022-10536-1}{\emph{Eur. Phys. J. C}
  {\bfseries 82} (2022) 869}
  [\href{https://arxiv.org/abs/2111.09849}{{\ttfamily 2111.09849}}].

\bibitem{Workman:2022ynf}
{\scshape Particle Data Group} collaboration, \emph{{Review of Particle
  Physics}}, \href{https://doi.org/10.1093/ptep/ptac097}{\emph{PTEP} {\bfseries
  2022} (2022) 083C01}.

\bibitem{Aebischer:2018bkb}
J.~Aebischer, J.~Kumar and D.M.~Straub, \emph{{Wilson: a Python package for the
  running and matching of Wilson coefficients above and below the electroweak
  scale}}, \href{https://doi.org/10.1140/epjc/s10052-018-6492-7}{\emph{Eur.
  Phys. J. C} {\bfseries 78} (2018) 1026}
  [\href{https://arxiv.org/abs/1804.05033}{{\ttfamily 1804.05033}}].

\bibitem{Leljak:2021vte}
D.~Leljak, B.~Meli\'c and D.~van Dyk, \emph{{The $ \overline{B} $
  \textrightarrow{} \ensuremath{\pi} form factors from QCD and their impact on
  $|V_{ub}|$}}, \href{https://doi.org/10.1007/JHEP07(2021)036}{\emph{JHEP}
  {\bfseries 07} (2021) 036}
  [\href{https://arxiv.org/abs/2102.07233}{{\ttfamily 2102.07233}}].

\bibitem{Gao:2019lta}
J.~Gao, C.-D.~L\"u, Y.-L.~Shen, Y.-M.~Wang and Y.-B.~Wei, \emph{{Precision
  calculations of $B \to V$ form factors from soft-collinear effective theory
  sum rules on the light-cone}},
  \href{https://doi.org/10.1103/PhysRevD.101.074035}{\emph{Phys. Rev. D}
  {\bfseries 101} (2020) 074035}
  [\href{https://arxiv.org/abs/1907.11092}{{\ttfamily 1907.11092}}].

\end{thebibliography}\endgroup

\end{document}